\documentclass[twocolumn]{aastex631}

\usepackage{amsmath}

\newcommand{\tet}{$\Theta^2$A\,Ori}
\newcommand{\hef}{$^{4}$He}
\newcommand{\het}{$^{3}$He}
\newcommand{\heiso}{\het/\hef}
\newcommand{\HeI}{He\,\textsc{i}}
\newcommand{\HeIs}{He\,\textsc{i}*}
\newcommand{\HeIw}{He\,\textsc{i}*\,$\lambda1.0833\,\mu$m}
\newcommand{\HeII}{He\,\textsc{ii}}
\newcommand{\HII}{H\,\textsc{ii}}
\newcommand{\obhh}{$\Omega_{\rm B,0}\,h^{2}$}
\newcommand{\neff}{$N_{\rm eff}$}
\newcommand{\vice}{\texttt{VICE}}

\shorttitle{The helium isotope ratio of the Orion Nebula}
\shortauthors{Cooke et al.}

\graphicspath{{./}{figures/}}

\begin{document}

\title{Primordial helium-3 redux: The helium isotope ratio of the Orion nebula\footnote{Based on observations collected at the European Organisation for Astronomical Research in the Southern Hemisphere under ESO programme(s) 107.22U1.001, 194.C-0833.}}

\author[0000-0001-7653-5827]{Ryan J. Cooke}
\affiliation{Centre for Extragalactic Astronomy, Durham University, Durham DH1 3LE, UK}
\email{ryan.j.cooke@durham.ac.uk}

\author[0000-0002-5777-1629]{Pasquier Noterdaeme}
\affiliation{Franco-Chilean Laboratory for Astronomy, IRL 3386, CNRS and U. de Chile, Casilla 36-D, Santiago, Chile}
\affiliation{Institut d'Astrophysique de Paris, UMR 7095, CNRS and SU, 98bis bd Arago, 75014 Paris, France}

\author[0000-0002-6534-8783]{James W. Johnson}
\affiliation{Department of Astronomy, The Ohio State University, 140 W. 18th Ave., Columbus OH 43210, USA}

\author[0000-0002-5139-4359]{Max Pettini}
\affiliation{Institute of Astronomy, University of Cambridge, Madingley Road, Cambridge, CB3 0HA, UK}

\author[0000-0002-0786-7307]{Louise Welsh}
\affiliation{Dipartimento di Fisica G. Occhialini, Universit\`a degli Studi di Milano Bicocca, Piazza della Scienza 3, 20126 Milano, Italy}
\affiliation{INAF – Osservatorio Astronomico di Brera, via Bianchi 46, I-23087 Merate (LC), Italy}

\author[0000-0002-4288-599X]{Celine Peroux}
\affiliation{European Southern Observatory, Karl-Schwarzschildstrasse 2, D-85748 Garching bei M\"unchen, Germany}
\affiliation{Aix Marseille Universit\'e, CNRS, LAM (Laboratoire d’Astrophysique de Marseille) UMR 7326, F-13388 Marseille, France}

\author[0000-0002-7040-5498]{Michael T. Murphy}
\affiliation{Centre for Astrophysics and Supercomputing, Swinburne University of Technology, Hawthorn, Victoria 3122, Australia}

\author{David H. Weinberg}
\affiliation{Department of Astronomy, The Ohio State University, 140 W. 18th Ave., Columbus OH 43210, USA}
\affiliation{Center for Cosmology and Astroparticle Physics, The Ohio State University, 191 W. Woodruff Ave., Columbus OH 43210, USA}



\begin{abstract}
We report the first direct measurement of the helium isotope ratio, \heiso, outside of the Local Interstellar Cloud, as part of science verification observations with the upgraded CRyogenic InfraRed Echelle Spectrograph (CRIRES). Our determination of \heiso\ is based on metastable \HeIs\ absorption along the line-of-sight towards \tet\ in the Orion Nebula. We measure a value
\heiso~$=(1.77\pm0.13)\times10^{-4}$,
which is just $\sim40$ per cent above the primordial relative abundance of these isotopes, assuming the Standard Model of particle physics and cosmology, (\het/\hef)$_{\rm p}=(1.257\pm0.017)\times10^{-4}$. We calculate a suite of galactic chemical evolution simulations to study the Galactic build up of these isotopes, using the yields from \citet{LimongiChieffi2018} for stars in the mass range $M=8-100~{\rm M}_{\odot}$ and \citet{Lagarde2011,Lagarde2012} for $M=0.8-8~{\rm M}_{\odot}$. We find that these simulations simultaneously reproduce the Orion and protosolar \heiso\ values if the calculations are initialized with a primordial ratio
$(^{3}{\rm He}/^{4}{\rm He})_{\rm p}=(1.043\pm0.089)\times10^{-4}$.
Even though the quoted error does not include the model uncertainty, this determination agrees with the Standard Model value to within $\sim2\sigma$.
We also use the present-day Galactic abundance of deuterium (D/H), helium (He/H), and \heiso\ to infer an empirical limit on the primordial \het\ abundance,
$(^{3}{\rm He/H})_{\rm p} \le (1.09\pm0.18)\times10^{-5}$, which also agrees with the Standard Model value.
We point out that it is becoming increasingly difficult to explain the discrepant primordial $^{7}$Li/H abundance with non-standard physics, without breaking the remarkable simultaneous agreement of three primordial element ratios (D/H, \hef/H, and \heiso) with the Standard Model values. 
\end{abstract}



\section{Introduction}
\label{sec:intro}

Most of the baryonic matter in the Universe was made within a few minutes following the Big Bang. This brief period of element genesis is commonly referred to as Big Bang Nucleosynthesis (BBN), and was primarily responsible for making the isotopes of the lightest chemical elements (for a review of this topic, see \citealt{Cyburt16,Pitrou18,Fields20}). The relative abundances of these elements are sensitive to the physical conditions and content of the Universe during the first few minutes. Thus, by measuring the relative abundances of the light elements made during BBN, we can learn about the physics of the very early Universe.

The abundances of the BBN nuclides are traditionally calculated and measured relative to the number of hydrogen atoms. Most studies have focused on determining the abundances of deuterium (D/H), helium-3 (\het/H), helium-4 (\hef/H), and lithium-7 ($^{7}$Li/H). Current measures of the primordial D/H and \hef/H abundances broadly agree with the values calculated assuming the Standard Model of particle physics and cosmology (\citealt{Izotov14,Cooke18,Fernandez19,Hsyu20,Aver21,Valerdi21,Kurichin2022}). However, the $^{7}$Li/H abundance inferred from the atmospheres of the most metal-poor halo stars deviates from the Standard Model value by $\sim 6\sigma$ (\citealt{Asplund06,Aoki09,Melendez10,Sbordone10,Pinto21}). At present, it is still unclear if the stellar atmospheres of metal-poor stars have burnt some of the lithium \citep{Korn06,Lind09}, or if an ingredient is missing from the Standard Model (for a comprehensive review on this topic, see \citealt{Fields11}).

The primordial abundance of \het\ has received relatively less attention, largely because it is so challenging to measure; for almost all of the helium atomic transitions, \hef\ swamps \het\ because it is $\sim10,000$ times more abundant and the isotope shifts of almost all transitions are $\lesssim15\,{\rm km~s}^{-1}$. However, because \het\ has a non-zero nuclear spin, the ground state of \het\ exhibits hyperfine structure, which gives rise to a \het$^{+}$ spin-flip transition at $8.7\,{\rm GHz}$ --- while \hef\ does not. This transition has enabled the only measurement of the \het/H abundance outside of the Solar System \citep{Bania02}, and has only been detected towards \HII\ regions in the Milky Way \citep{BalBan18}.\footnote{There are also some reported detections of the $^{3}$He$^{+}$~8.7~GHz line from a small number of planetary nebulae \citep{Balser99,Balser06,GuzmanRamirez16}; however, these detections have not been confirmed with more recent data \citep{BanBal21}.} Based on these observations, there appears to be a gentle decrease of \het/H with increasing galactocentric radius, in line with models of galactic chemical evolution \citep{Lagarde2012}. The best available estimate of the \het/H ratio of the outer Milky Way is \het/H~$=(1.10\pm0.28)\times10^{-5}$, which is solely based on the most distant and well-characterized \HII\ region (S209) where the He$^{+}$~8.7~GHz line has been detected \citep{BalBan18}. Unfortunately, measurements of the Milky Way \het\ abundance may not represent the primordial value due to post-BBN production of \het; future measurements of this isotope in near-pristine environments are required to understand the complete cosmic chemical evolution of \het, and to secure its primordial abundance.

While the abundance of \het\ is challenging to measure, there are several proposed approaches that may secure a measurement of the primordial \het\ abundance with future facilities. Akin to the detection of \het$^{+}$ $8.7~{\rm GHz}$ in Galactic \HII\ regions, it may be possible (but extremely challenging) to detect \het$^{+}$ emission around growing supermassive black holes at redshift $z>12$ \citep{Vasiliev19}, provided that the quasar environment retains a primordial composition. Another possibility is to detect the \het$^{+}$ $8.7~{\rm GHz}$ transition in absorption against the spectrum of a radio bright quasar during helium reionization (at redshift $z\sim3$; \citealt{McQSwi2009,Takeuchi14,Khullar20}). One advantage of this approach is that the gas in the intergalactic medium is largely unprocessed, and the \het/H value would therefore closely reflect the primordial value. The intergalactic medium is also somewhat simpler to model than the Galactic \HII\ regions where this \het$^{+}$ transition has previously been studied in emission.

An alternative approach proposed by \citet{Cooke15} is to use a combination of optical and near-infrared transitions to measure the helium isotope ratio (\heiso). Because the ionization potential of \het\ is almost identical to that of \hef, the helium isotope ratio is much less sensitive to ionization corrections than \het/H. Furthermore, \citet{Cooke15} highlighted that the helium isotope ratio and D/H provide orthogonal bounds on the present day baryon density (\obhh) and the effective number of neutrino species (\neff). Finally, the isotope shifts of the optical and near-infrared transitions are all different, so a relative comparison of any two line profiles would allow one to unambiguously identify \het.

The only available measures of the helium isotope ratio are based on terrestrial and Solar System environments, including the Earth's mantle (\heiso~$=[1.1-4.2]\times10^{-5}$; \citealt{Peron18}), meteorites (\heiso~$=[1.47-190]\times10^{-4}$; \citealt{Buse00,Kri21}), the Local Interstellar Cloud (\heiso~$=1.62\pm0.29\times10^{-4}$; \citealt{Buse06}), Jupiter (\heiso~$=1.66\pm0.05\times10^{-4}$; \citealt{Mah1998}), and solar wind particles (\heiso~$=[4.5-4.8]\times10^{-4}$; \citealt{Heber2012}). The elevated values measured from the present day solar wind reflect the burning of deuterium into \het\ during the pre-main-sequence evolution of the Sun. Therefore, the \heiso\ measure based on Jupiter is our best estimate of the proto-solar value of the helium isotope ratio.

In this paper we propose a new approach, qualitatively similar to that described by \citet{Cooke15}, to measure the helium isotope ratio. Using this approach, we report the first direct measurement of \het/\hef\ beyond the Local Interstellar Cloud. Our approach uses the light of a background object (in our case, a bright O star) to study the absorbing material along the line-of-sight. Since all ground state (singlet) transitions of neutral helium are in the far ultraviolet (see e.g. \citealt{CooFum18}), the only helium absorption lines accessible to ground-based telescopes arise from excited states of metastable helium (\HeIs), which has a lifetime of $\sim130~$min.

Interstellar absorption lines due to metastable helium were first identified toward the Orion Nebula nearly a century ago \citep{Wilson37}. They have since been detected towards five stars associated with the Orion Nebula \citep{Odell93,Oud97}, towards several stars in the young open cluster NGC\,6611 \citep{Evans05}, and towards $\zeta$~Oph \citep{GalKre12}. Extragalactic \HeIs\ absorption lines have also been detected in the host galaxy of the gamma-ray burst GRB\,140506A at redshift $z=0.889$ \citep{Fynbo14}. Several broad absorption line quasars also exhibit \HeIs\ absorption, although the kinematics of these features are significantly broader ($\sim1000\,{\rm km~s}^{-1}$; \citealt{Liu2015}) and more complicated than the interstellar features ($\sim10\,{\rm km~s}^{-1}$). The interstellar \HeIs\ absorption profiles tend to be quiescent and smooth, suggesting that a very simple broadening mechanism is responsible for the line shape, and the absorbing gas is homogeneous. Furthermore, given the lifetime of the metastable state and the fact that metastable helium only occurs in regions where He$^{+}$ is recombining, the \HeIs\ absorption likely occurs at the edge of the \HeII\ ionization region around the hottest stars.

The simplicity and quiescence ($\lesssim10~{\rm km~s}^{-1}$) of the line profiles is one of the key benefits of using \HeIs\ absorption to measure the helium isotope ratio. Demonstrating the promise of this new approach for measuring the primordial \heiso\ ratio is the primary motivation of this work. In Section~\ref{sec:obs} we provide the details of our CRIRES Science Verification observations. The atomic data are described in Section~\ref{sec:atomic} and our absorption line profile analysis is presented in Section~\ref{sec:analysis}. We discuss the implications of our measurement in Section~\ref{sec:results} before summarizing our conclusions in Section~\ref{sec:conc}.

\section{Observations}
\label{sec:obs}

Motivated by the strong, kinematically quiescent \HeIs\ absorption towards \tet\ \citep{Odell93}, we obtained Very Large Telescope (VLT) CRyogenic InfraRed Echelle Spectrograph (CRIRES; \citealt{CRIRES2004}) observations of \tet\ as part of the CRIRES Upgrade Project \citep{CRIRESp} science verification on 2021 Sep 18. CRIRES is an infrared cross-dispersed echelle spectrograph covering a wavelength range from 0.95 to 5.3\,$\mu$m at a spectral resolving power $R\simeq40,000$ or $R\simeq80,000$. As part of the upgrade, a cross-disperser was installed to increase the simultaneously covered wavelength range by a factor of ten, and three sensitive Hawaii 2RG detectors were installed. CRIRES is based at the Nasmyth focus of UT3, and is fed by the Multi-Applications Curvature Adaptive Optics (MACAO) system.

\subsection{CRIRES data}

The goal of our observations is to detect the $^{3}$\HeIs\,$\lambda1.0833\,\mu$m absorption line to obtain the first direct measure of the helium isotope ratio beyond the Local Interstellar Cloud. Detecting $^{3}$\HeIs\ absorption is challenging, as we expect the $^{3}$\HeIs\ absorption line to be very weak, and only detectable in spectra of both high resolution ($R\gtrsim40,000$) and very high signal-to-noise ratio (${\rm S/N}>1000$). To minimise the effect of pixel-to-pixel sensitivity variations affecting the final combined S/N, we designed the observing programme to ensure that the target spectrum was projected onto different detector pixels with each exposure. A total of ten exposures were acquired at two nodding positions (20 exposures total); each nod was separated by $7.5''$, and the target spectrum was shifted by $0.5''$ on the detector with each exposure. To minimize the effects of non-linearity, the detector integration time was \texttt{DIT}=5\,s, and each exposure contains \texttt{NDIT}=20 detector integrations, resulting in an exposure time of 100~s. We also took four shorter exposures; two with \texttt{NDIT}=3 and two with \texttt{NDIT}=1. Therefore, the total integration time on source was 2040\,s. We used the $0.2''$ slit in combination with the Multi-Applications Curvature Adaptive Optics (MACAO) system. The nominal spectral resolving power of this setup is $R\sim80,000$.

\begin{figure}
\includegraphics[width=\columnwidth]{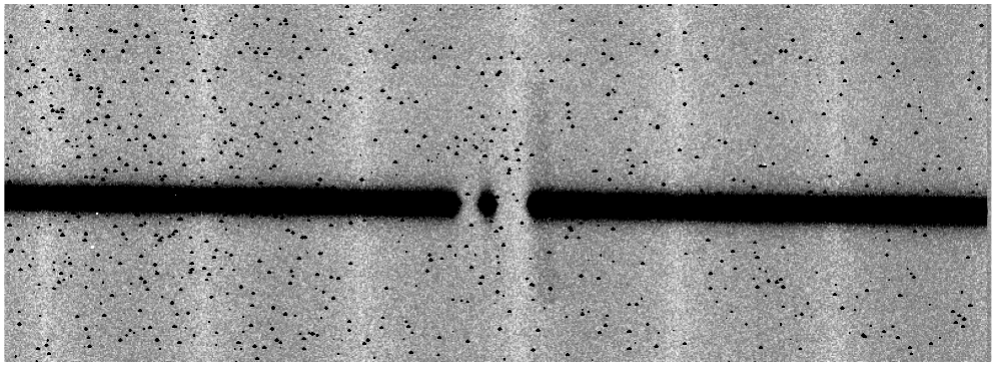}

\vspace{0.1cm}

\includegraphics[width=\columnwidth]{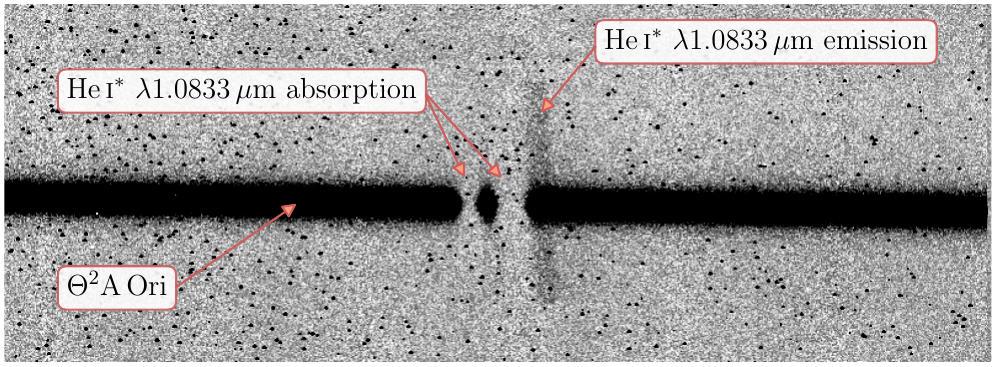}

\vspace{0.1cm}

\includegraphics[width=\columnwidth]{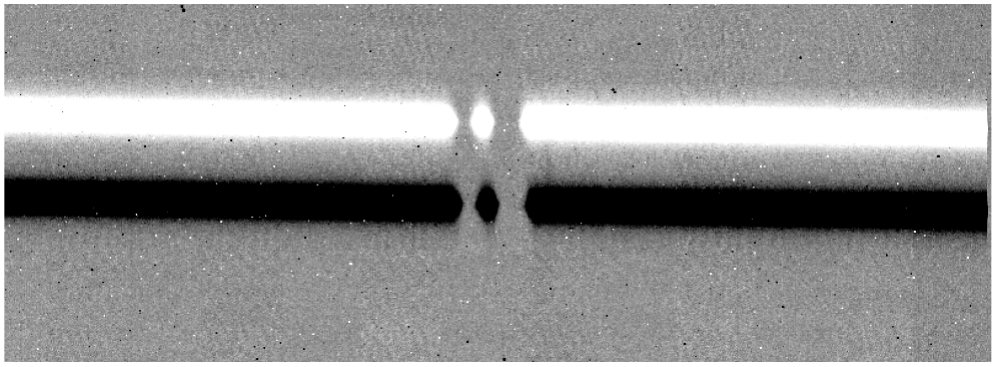}
\caption{An example of a raw CRIRES frame (top), after detector pattern removal (middle), and after differencing a pair of science frames at different nod positions (bottom). These panels show the \HeIs\ absorption superimposed on the spectrum of \tet. Wavelength increases from left to right. Faint \HeIs\ emission is detected in the red wing of the line profile (see middle panel). Since the emission only appears in the red wing, it presumably arises at the edge of an expanding shell. The `pepper' that can be seen in the top two frames is due to hot pixels, that are largely removed as part of differencing (see bottom panel). \label{fig:2draw}}
\end{figure}

The data were reduced using a set of custom-built routines in combination with the PypeIt data reduction pipeline \citep{PypeIt}.\footnote{For installation and examples, see \url{https://pypeit.readthedocs.io/en/latest/}}
A combination of nine flatfield frames were used to remove the pixel-to-pixel sensitivity variations. We subtracted a pair of science frames at different nod positions to remove the periodic CRIRES detector pattern, hot pixels, and  any background emission (see Figure~\ref{fig:2draw}).\footnote{Since the sky background is insignificant compared to the brightness of the source, we also reduced the individual raw frames, and found this resulted in a lower final S/N ratio by a factor of $\sim2$. While we were unable to unequivocally identify the source of this reduced S/N, it is presumably related to either the persistent detector pattern noise that is imprinted on the CRIRES detector or unidentified `hot' pixels that are otherwise removed during differencing. The pattern noise is periodic along the spectral dimension, and appears to be constant for all exposures (see top panel of Figure~\ref{fig:2draw} for an example). By subtracting science frames at different nod positions, we were able to fully remove the detector pattern and hot pixels in the difference frames (see bottom panel of Figure~\ref{fig:2draw}).} Bad pixels were identified and masked, and the object was traced and optimally extracted using PypeIt. A spectrum of the background emission (i.e. the sky and \HII\ region emission) was also extracted from the individual science frames (i.e. without subtracting observations at different nod positions), and used for validation (see Section~\ref{sec:validate}).

\label{sec:fitting}

The wavelength calibration was performed using a second order polynomial fit to the telluric absorption lines imprinted on the target spectrum (see \citealt{Seifahrt2010}). Due to the low line density (six telluric lines spanning 2048 pixels) and the weakness of these features, the solution was not sufficiently accurate given the high S/N desired. We therefore used the telluric solution as a first guess, and performed a simultaneous fit to all individual exposures to construct a model of the continuum, absorption, and wavelength solution (see Section~\ref{sec:fitting} for details of the line fitting procedure). The wavelength correction comprised a simple shift and stretch to the telluric wavelength solution, using as a reference archival optical observations of \HeIs\,$\lambda3188$\,\AA\ (described in Section~\ref{sec:uves}). The continuum was modelled with a global high order Legendre polynomial for all exposures, combined with a multiplicative scale and tilt to account for the relative sensitivity of each exposure. Deviant pixels were masked during the fitting. We applied the wavelength and continuum corrections to each extracted spectrum, resampled all exposures onto a single wavelength scale,\footnote{We used the \textsc{linetools} package \textsc{XSpectrum1d}, to ensure that flux is conserved. \textsc{linetools} is available from \url{https://linetools.readthedocs.io/en/latest/}} and combined all exposures while sigma-clipping (a rejection threshold of $3\sigma$).

\begin{figure}
\includegraphics[width=\columnwidth]{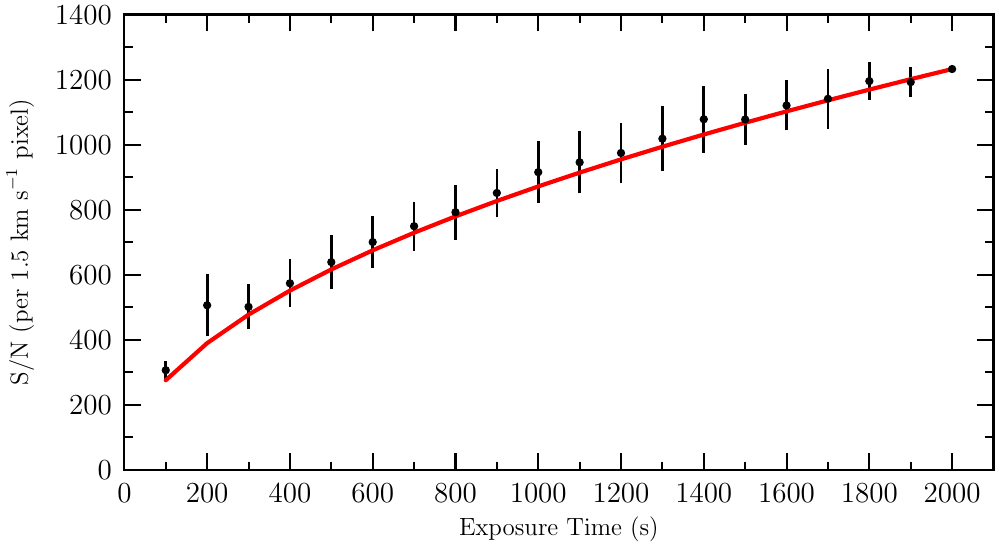}
\caption{The increase of the final combined S/N as the exposure time is increased. The black points with error bars are based on randomly combining exposures of a set total exposure time, where the error bar represents the dispersion. The red curve shows the expected relation, where the signal-to-noise ratio is proportional to the square root of the exposure time. Note that each data point is not independent.\label{fig:snrexp}}
\end{figure}

The inferred error spectrum is dominated by the photon count uncertainty of the target spectrum. However, this does not include additional noise terms that arise due to instrument and data reduction systematics (e.g. residual pixel-to-pixel sensitivity variations, wavelength/continuum calibration, etc.). To uniformly account for unknown systematic uncertainties, we fit a low order Legendre polynomial to the target continuum, calculate the observed deviations about this model, and scale the error spectrum so that it represents more faithfully the observed fluctuations of the data. From these fluctuations, we measure a S/N=1260 per $1.5~{\rm km~s}^{-1}$ pixel, based on the blue-side continuum near the \HeIs\,$\lambda1.0833\,\mu$m absorption line.\footnote{The achieved S/N is almost a factor of $\sim2$ lower than that expected for the requested integration time (S/N~$\gtrsim2200$). The reduced S/N is likely a combination of residual data reduction systematics (e.g. unaccounted for pixel-to-pixel sensitivity variations) and the poor centering of the target in the slit due to an acquisition error (private communication, ESO User Support Department).} To test if the S/N we achieve is limited by data reduction systematics, we combined a random subset of $n$ exposures to determine how the S/N increases as the total exposure time increases. We repeated this process 100 times for $n=1,2\ldots20$; the mean values and dispersions for each $n$ (recall, each exposure time is 100\,s) are shown as the black points with error bars in Figure~\ref{fig:snrexp}. The expected growth of the S/N, assuming it increases as the square root of the exposure time, is shown as the red curve. The good agreement between the data and the expected growth suggests that the S/N is currently limited by exposure time, and not data reduction systematics.

\subsection{UVES data}
\label{sec:uves}
We supplemented our CRIRES observations of \tet\ with archival VLT Ultraviolet and Visual Echelle Spectrograph (UVES) observations of this target, acquired on 2014 September 24 (Programme ID: 194.C-0833).\footnote{The data can be accessed from the UVES Science Portal, available from: \url{http://archive.eso.org/scienceportal/home}} These data were acquired as part of the ESO Diffuse Interstellar Bands Large Exploration Survey (EDIBLES); for further details about the instrument setup and data reduction, see \citet{EDIBLES}. The nominal spectral resolution of these data are $R\simeq71050$, and cover the \HeIs\ transitions at $3188\,$\AA\ and $3889\,$\AA.

\section{Atomic Data}\label{sec:atomic}

\begin{table*}
    \centering
    \caption{Atomic data of the \HeIs\ triplet transitions used in this work.}
    \label{tab:atomic}
    \begin{tabular}{lcccccccc}
        \hline
        Ion & Lower State & Upper State & $F$ & $F'$ & $\lambda_{0}$ & $f$ & $\gamma_{ul}$ & $\Delta v_{3-4}$ \\
            &             &             &     &      &     (\AA)     &     & $(10^{7}~{\rm s}^{-1})$ & (km~s$^{-1}$) \\
        \hline
 $^{3}$He\,\textsc{i}$^{*}$ & 1s2s\,$^{3}{\rm S}_{1}$ & 1s2p\,$^{3}{\rm P}^{\rm o}_{0}$ & 1.5 & 0.5 & 10833.27471 & 0.03996 & 1.022 & +33.7 \\
 $^{3}$He\,\textsc{i}$^{*}$ & 1s2s\,$^{3}{\rm S}_{1}$ & 1s2p\,$^{3}{\rm P}^{\rm o}_{0}$ & 0.5 & 0.5 & 10833.53856 & 0.01998 & 1.022 & +41.0 \\
 $^{3}$He\,\textsc{i}$^{*}$ & 1s2s\,$^{3}{\rm S}_{1}$ & 1s2p\,$^{3}{\rm P}^{\rm o}_{1}$ & 1.5 & 1.5 & 10834.55125 & 0.09989 & 1.022 & +36.9 \\
 $^{3}$He\,\textsc{i}$^{*}$ & 1s2s\,$^{3}{\rm S}_{1}$ & 1s2p\,$^{3}{\rm P}^{\rm o}_{1}$ & 0.5 & 1.5 & 10834.81516 & 0.01998 & 1.022 & +44.2 \\
 $^{3}$He\,\textsc{i}$^{*}$ & 1s2s\,$^{3}{\rm S}_{1}$ & 1s2p\,$^{3}{\rm P}^{\rm o}_{1}$ & 1.5 & 0.5 & 10834.37457 & 0.01998 & 1.022 & +32.0 \\
 $^{3}$He\,\textsc{i}$^{*}$ & 1s2s\,$^{3}{\rm S}_{1}$ & 1s2p\,$^{3}{\rm P}^{\rm o}_{1}$ & 0.5 & 0.5 & 10834.63847 & 0.03996 & 1.022 & +39.3 \\
 $^{3}$He\,\textsc{i}$^{*}$ & 1s2s\,$^{3}{\rm S}_{1}$ & 1s2p\,$^{3}{\rm P}^{\rm o}_{2}$ & 1.5 & 2.5 & 10834.62099 & 0.17980 & 1.022 & +36.4 \\
 $^{3}$He\,\textsc{i}$^{*}$ & 1s2s\,$^{3}{\rm S}_{1}$ & 1s2p\,$^{3}{\rm P}^{\rm o}_{2}$ & 1.5 & 1.5 & 10834.34842 & 0.01998 & 1.022 & +28.8 \\
 $^{3}$He\,\textsc{i}$^{*}$ & 1s2s\,$^{3}{\rm S}_{1}$ & 1s2p\,$^{3}{\rm P}^{\rm o}_{2}$ & 0.5 & 1.5 & 10834.61232 & 0.09989 & 1.022 & +36.1 \\
 $^{3}$He\,\textsc{i}$^{*}$ & 1s2s\,$^{3}{\rm S}_{1}$ & 1s3p\,$^{3}{\rm P}^{\rm o}_{0}$ & 1.5 & 0.5 & 3889.90165 & 0.00478 & 1.055 & +15.0 \\
 $^{3}$He\,\textsc{i}$^{*}$ & 1s2s\,$^{3}{\rm S}_{1}$ & 1s3p\,$^{3}{\rm P}^{\rm o}_{0}$ & 0.5 & 0.5 & 3889.93567 & 0.00239 & 1.055 & +17.7 \\
 $^{3}$He\,\textsc{i}$^{*}$ & 1s2s\,$^{3}{\rm S}_{1}$ & 1s3p\,$^{3}{\rm P}^{\rm o}_{1}$ & 1.5 & 1.5 & 3889.93000 & 0.01194 & 1.055 & +14.1 \\
 $^{3}$He\,\textsc{i}$^{*}$ & 1s2s\,$^{3}{\rm S}_{1}$ & 1s3p\,$^{3}{\rm P}^{\rm o}_{1}$ & 0.5 & 1.5 & 3889.96402 & 0.00239 & 1.055 & +16.7 \\
 $^{3}$He\,\textsc{i}$^{*}$ & 1s2s\,$^{3}{\rm S}_{1}$ & 1s3p\,$^{3}{\rm P}^{\rm o}_{1}$ & 1.5 & 0.5 & 3889.94461 & 0.00239 & 1.055 & +15.2 \\
 $^{3}$He\,\textsc{i}$^{*}$ & 1s2s\,$^{3}{\rm S}_{1}$ & 1s3p\,$^{3}{\rm P}^{\rm o}_{1}$ & 0.5 & 0.5 & 3889.97863 & 0.00478 & 1.055 & +17.8 \\
 $^{3}$He\,\textsc{i}$^{*}$ & 1s2s\,$^{3}{\rm S}_{1}$ & 1s3p\,$^{3}{\rm P}^{\rm o}_{2}$ & 1.5 & 2.5 & 3889.96332 & 0.02149 & 1.055 & +16.4 \\
 $^{3}$He\,\textsc{i}$^{*}$ & 1s2s\,$^{3}{\rm S}_{1}$ & 1s3p\,$^{3}{\rm P}^{\rm o}_{2}$ & 1.5 & 1.5 & 3889.96058 & 0.00239 & 1.055 & +16.2 \\
 $^{3}$He\,\textsc{i}$^{*}$ & 1s2s\,$^{3}{\rm S}_{1}$ & 1s3p\,$^{3}{\rm P}^{\rm o}_{2}$ & 0.5 & 1.5 & 3889.99460 & 0.01194 & 1.055 & +18.8 \\
 $^{3}$He\,\textsc{i}$^{*}$ & 1s2s\,$^{3}{\rm S}_{1}$ & 1s4p\,$^{3}{\rm P}^{\rm o}_{0}$ & 1.5 & 0.5 & 3188.79672 & 0.00191 & 0.657 & +13.3 \\
 $^{3}$He\,\textsc{i}$^{*}$ & 1s2s\,$^{3}{\rm S}_{1}$ & 1s4p\,$^{3}{\rm P}^{\rm o}_{0}$ & 0.5 & 0.5 & 3188.81958 & 0.00095 & 0.657 & +15.5 \\
 $^{3}$He\,\textsc{i}$^{*}$ & 1s2s\,$^{3}{\rm S}_{1}$ & 1s4p\,$^{3}{\rm P}^{\rm o}_{1}$ & 1.5 & 1.5 & 3188.80263 & 0.00477 & 0.657 & +12.8 \\
 $^{3}$He\,\textsc{i}$^{*}$ & 1s2s\,$^{3}{\rm S}_{1}$ & 1s4p\,$^{3}{\rm P}^{\rm o}_{1}$ & 0.5 & 1.5 & 3188.82549 & 0.00095 & 0.657 & +15.0 \\
 $^{3}$He\,\textsc{i}$^{*}$ & 1s2s\,$^{3}{\rm S}_{1}$ & 1s4p\,$^{3}{\rm P}^{\rm o}_{1}$ & 1.5 & 0.5 & 3188.81781 & 0.00095 & 0.657 & +14.3 \\
 $^{3}$He\,\textsc{i}$^{*}$ & 1s2s\,$^{3}{\rm S}_{1}$ & 1s4p\,$^{3}{\rm P}^{\rm o}_{1}$ & 0.5 & 0.5 & 3188.84067 & 0.00191 & 0.657 & +16.4 \\
 $^{3}$He\,\textsc{i}$^{*}$ & 1s2s\,$^{3}{\rm S}_{1}$ & 1s4p\,$^{3}{\rm P}^{\rm o}_{2}$ & 1.5 & 2.5 & 3188.82480 & 0.00859 & 0.657 & +14.8 \\
 $^{3}$He\,\textsc{i}$^{*}$ & 1s2s\,$^{3}{\rm S}_{1}$ & 1s4p\,$^{3}{\rm P}^{\rm o}_{2}$ & 1.5 & 1.5 & 3188.82404 & 0.00095 & 0.657 & +14.8 \\
 $^{3}$He\,\textsc{i}$^{*}$ & 1s2s\,$^{3}{\rm S}_{1}$ & 1s4p\,$^{3}{\rm P}^{\rm o}_{2}$ & 0.5 & 1.5 & 3188.84690 & 0.00477 & 0.657 & +16.9 \\
 $^{4}$He\,\textsc{i}$^{*}$ & 1s2s\,$^{3}{\rm S}_{1}$ & 1s2p\,$^{3}{\rm P}^{\rm o}_{0}$ & $\ldots$ & $\ldots$ & 10832.05747 & 0.05993 & 1.022 & $\ldots$ \\
 $^{4}$He\,\textsc{i}$^{*}$ & 1s2s\,$^{3}{\rm S}_{1}$ & 1s2p\,$^{3}{\rm P}^{\rm o}_{1}$ & $\ldots$ & $\ldots$ & 10833.21675 & 0.17980 & 1.022 & $\ldots$ \\
 $^{4}$He\,\textsc{i}$^{*}$ & 1s2s\,$^{3}{\rm S}_{1}$ & 1s2p\,$^{3}{\rm P}^{\rm o}_{2}$ & $\ldots$ & $\ldots$ & 10833.30644 & 0.29967 & 1.022 & $\ldots$ \\
 $^{4}$He\,\textsc{i}$^{*}$ & 1s2s\,$^{3}{\rm S}_{1}$ & 1s3p\,$^{3}{\rm P}^{\rm o}_{0}$ & $\ldots$ & $\ldots$ & 3889.70656 & 0.00716 & 1.055 & $\ldots$ \\
 $^{4}$He\,\textsc{i}$^{*}$ & 1s2s\,$^{3}{\rm S}_{1}$ & 1s3p\,$^{3}{\rm P}^{\rm o}_{1}$ & $\ldots$ & $\ldots$ & 3889.74751 & 0.02149 & 1.055 & $\ldots$ \\
 $^{4}$He\,\textsc{i}$^{*}$ & 1s2s\,$^{3}{\rm S}_{1}$ & 1s3p\,$^{3}{\rm P}^{\rm o}_{2}$ & $\ldots$ & $\ldots$ & 3889.75083 & 0.03582 & 1.055 & $\ldots$ \\
 $^{4}$He\,\textsc{i}$^{*}$ & 1s2s\,$^{3}{\rm S}_{1}$ & 1s4p\,$^{3}{\rm P}^{\rm o}_{0}$ & $\ldots$ & $\ldots$ & 3188.65492 & 0.00286 & 0.657 & $\ldots$ \\
 $^{4}$He\,\textsc{i}$^{*}$ & 1s2s\,$^{3}{\rm S}_{1}$ & 1s4p\,$^{3}{\rm P}^{\rm o}_{1}$ & $\ldots$ & $\ldots$ & 3188.66614 & 0.00859 & 0.657 & $\ldots$ \\
 $^{4}$He\,\textsc{i}$^{*}$ & 1s2s\,$^{3}{\rm S}_{1}$ & 1s4p\,$^{3}{\rm P}^{\rm o}_{2}$ & $\ldots$ & $\ldots$ & 3188.66706 & 0.01432 & 0.657 & $\ldots$ \\
        \hline
    \end{tabular}

    From left to right, the columns represent:
    (1) The ion;
    (2/3) The lower/upper state of the transition;
    (4 and 5) The total rotational quantum number including nuclear spin of the ground state ($J=1$) of \het;
    (6) Vacuum wavelength of the transition;
    (7) Oscillator strength of the transition;
    (8) Natural damping constant of the transition;
    (9) Isotope shift of \het\ relative to \hef.
\end{table*}

\begin{figure*}
\centering
\includegraphics[width=0.8\textwidth]{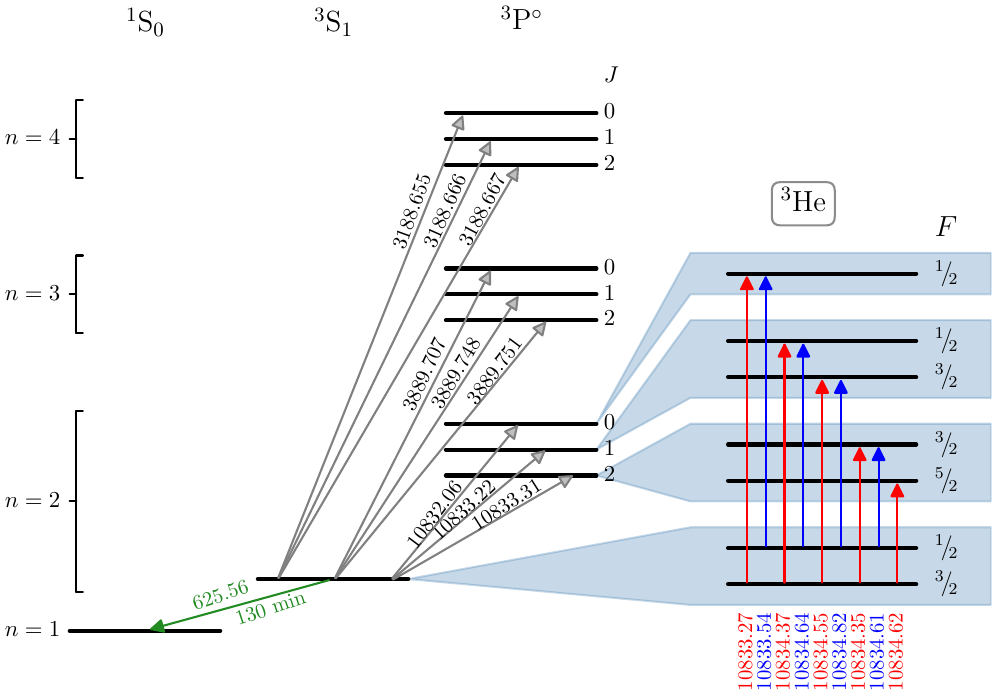}
\caption{Grotrian diagram (not to scale) showing the energy levels of the lowest levels of the helium triplet state. All $^{4}$\HeIs\ transitions used in this work are shown in gray. The hyperfine structure of the $n=2$ level of \het\ is shown in the blue shaded boxes. The blue and red arrows indicate transitions from the $F=0.5$ and $F=1.5$ levels, respectively. The decay of metastable helium to the singlet state is represented by a green arrow. All wavelengths are shown in units of \AA.\label{fig:grotrian}}
\end{figure*}

The helium absorption lines analyzed in this paper originate from the metastable $2^{3}$S state of \HeI, which has a lifetime of $\sim130~{\rm min}$. This state is populated by recombinations from the \HeII\ state, when the recombining electron has the same spin as the electron in the \HeII\ ground state. The excited-state\footnote{We denote the excited-state with an apostrophe.} of the absorption lines (the $^{3}{\rm P}^{\rm o}$ state) exhibits fine-structure, with total angular momentum quantum numbers $J'=0,1,2$, while the ground state of metastable helium (the $^{3}{\rm S}$ state) has $J=1$.

We use the energy levels compiled by \citet{MorWuDra06} to determine the vacuum wavelengths of the helium transitions used in this paper. The oscillator strengths ($f$) for the \hef\ transitions were retrieved from the National Institutes of Standards and Technology (NIST) Atomic Spectral Database (ASD) \citep{NIST_ASD}. These oscillator strengths were also used for the \het\ transitions, however, because \het\ has a nuclear spin $I=1/2$, each fine-structure level with $J>0$ is split into two levels with $F=J\pm1/2$. In some cases, this hyperfine splitting is comparable to the fine-structure splitting, and may affect the shape of the line profile. To account for this, we calculate the relative probability of the transitions to the excited hyperfine levels as the product of the level degeneracies and the Wigner $6J$-symbol (see e.g. \citealt{MurBer2014}):\footnote{To calculate the Wigner $6J$-symbol, we used \textsc{SymPy} \citep{SymPy}, which is available from: \url{https://www.sympy.org/en/index.html}}
\begin{equation}
    S=(2F+1)(2F'+1)
    \begin{Bmatrix}
    J  & F  & I \\
    F' & J' & 1
    \end{Bmatrix}^{2}
\end{equation}
Finally, the natural damping constant ($\gamma_{ul}$) of each transition was computed by summing the spontaneous transition probabilities ($A_{ul}$; retrieved from \citealt{NIST_ASD}) to all lower levels,
\begin{equation}
    \gamma_{ul} = \sum_l A_{ul}
\end{equation}
All of the atomic data are compiled in Table~\ref{tab:atomic}, and a Grotrian diagram of the most relevant transitions is shown in Figure~\ref{fig:grotrian}. We also list the \het\ isotope shift of each transition in the final column of Table~\ref{tab:atomic}. Note that the isotope shift is largest for the \HeIs\,$\lambda1.0833\,\mu$m line, with an $f$-weighted isotope shift of $\sim +36.6~{\rm km~s}^{-1}$. Also note that the typical isotope shift is different for the 2s$\rightarrow$2p, 2s$\rightarrow$3p, and 2s$\rightarrow$4p, transitions; by comparing the profiles of \HeIs\,$\lambda3889$\AA\ and \HeIs\,$\lambda1.0833\,\mu$m, we can unambiguously determine if a \HeIs\,$\lambda1.0833\,\mu$m absorption feature is due to \het, or if it is due to coincident \hef\ absorption located at $\sim +36.6~{\rm km~s}^{-1}$ relative to the dominant absorption component (see Section~\ref{sec:validate}).

Lastly, throughout this work we assume that the excitation fractions of $^{3}$\HeIs\ and $^{4}$\HeIs\ are identical, so that the intrinsic helium isotope ratio is simply \heiso\,=\,$N(^{3}$\HeIs)/$N(^{4}$\HeIs). Given the similar ionization potential of the helium isotopes, we expect charge transfer reactions between \het\ and \hef\ to ensure this assumption is a reliable one. However, we note that this assumption may need to be considered in more detail when the precision of this measurement improves.

\section{Analysis}\label{sec:analysis}

Our new CRIRES data reveal that the strength and structure of the $^{4}$\HeIw\ absorption profile is quiescent and qualitatively similar to the expected profile based on the optical data that have been acquired over the past $\sim30$ years \citep{Odell93,EDIBLES}. This suggests that \HeIs\ is approximately in ionization equilibrium, since the lifetime of the metastable state ($\sim130~{\rm min}$) is significantly less than the time between the observations. Furthermore, at the expected location of the $^{3}$\HeIs\ absorption, we detect a significant absorption feature ($>13\sigma$) with a rest frame equivalent width of $EW=3.64\pm0.22$~m\AA\ (see top right panel of Figure~\ref{fig:modelfits}). Note, this $EW$ measure is a blend of the transitions to the $J'=1,2$ levels, so the effective oscillator strength of this absorption feature is $f_{\rm 1,2}=0.4794$ (i.e. the sum of all $f$ values in Table~\ref{tab:atomic} with an upper state 1s2p\,$^{3}{\rm P}^{\rm o}_{1}$ and 1s2p\,$^{3}{\rm P}^{\rm o}_{2}$). Given that this feature is extremely weak, we can estimate the corresponding $^3$\HeIs\ column density towards \tet\ using the relation \citep{Spitzer1978}:
\begin{eqnarray}
        \log_{10}\,N(^{3}{\rm He\,\textsc{i}^{*}})/{\rm cm}^{-2}&=&20.053+\log_{10}\bigg(\frac{EW}{\lambda_{0}^{2}\,f_{\rm 1,2}}\bigg)\nonumber\\
        &=&9.863\pm0.027\label{eqn:ewcoldens}
\end{eqnarray}
where $EW$ and $\lambda_{0}$ are in units of \AA\ and $\lambda_{0}\simeq10834.6$\,\AA\ is the approximate rest-frame wavelength of \het. In the following subsection, we perform a detailed profile analysis to determine the \heiso\ ratio along the line-of-sight to \tet.

\subsection{Profile fitting}
To model the \HeIs\ absorption lines, we use the Absorption LIne Software (\textsc{alis}) package.\footnote{\textsc{alis} is available for download from \url{https://github.com/rcooke-ast/ALIS}} \textsc{alis} uses a Levenberg-Marquardt algorithm to determine the model parameters that best-fit the data, by minimizing the chi-squared statistic. We perform a simultaneous fit to the stellar continuum, the interstellar absorption, the wavelength calibration, the zero-level, telluric absorption, and the instrument resolution. This fitting approach allows us to propagate the uncertainties of these quantities to the model parameters.

Before describing the implementation of this fitting procedure in detail, we first draw attention to a faint emission feature that is detected in the red wing ($\sim +23~{\rm km~s}^{-1}$) of the \HeIs$\,\lambda1.0833\,\mu$m profile (see Figure~\ref{fig:2draw}). While this emission is largely removed as part of the differencing process (see Section~\ref{sec:obs}), we note that the absorbing medium may be relatively thin owing to the conditions required for \HeIs\ absorption. As a result, some \HeIs\ emission may be superimposed on the \HeIs\ absorption, and this may become particularly pronounced in the wings of the profile as the absorption becomes optically thin. We therefore model the zero-level of this line with two Gaussian profiles in addition to a constant offset. The centroids of these two Gaussians have a fixed separation of 1.216~\AA, which corresponds to the wavelength difference between the $J'=0$ and $J'=1,2$ levels. We note that this choice does not impact the weak $^{3}$\HeIs\ absorption line profile (see Section~\ref{sec:validate}).

We model the stellar continuum as a high order Legendre polynomial, and the instrument resolution is assumed to be well-described by a Gaussian profile. The wavelength calibration is assumed to be correct for the \HeIs\,$\lambda3188$\,\AA\ line (since it is the weakest \HeIs\ line detected), and we include model parameters to apply a small shift and stretch correction to the wavelength scales of the \HeIs\,$\lambda3889$\,\AA\ and $\lambda1.0833\,\mu$m lines. We fit directly to the total column density of $^4$\HeIs\ and the column density ratio, $N(^{3}$\HeIs)/$N(^{4}$\HeIs). As described in Section~\ref{sec:atomic}, we assume that charge transfer reactions ensure that the intrinsic helium isotope ratio \heiso~=~$N(^{3}$\HeIs)/$N(^{4}$\HeIs).

\begin{figure*}
\includegraphics[width=\textwidth]{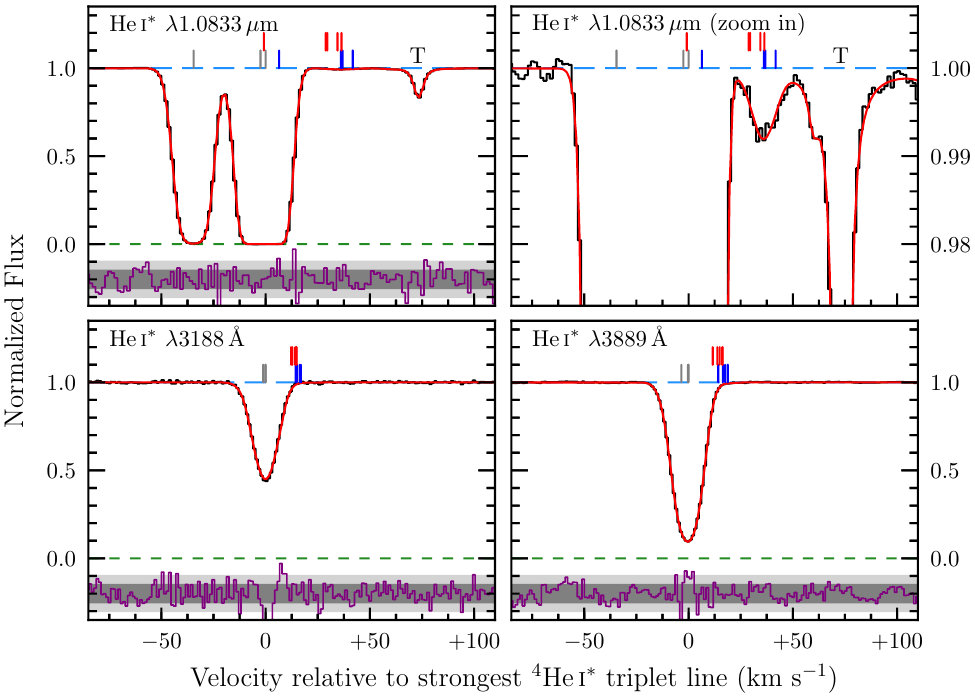}
\caption{Spectra of a \HeIs\ absorption line system towards \tet\ (top and bottom panels show the CRIRES and UVES data, respectively). The best-fitting model (red curves) are overlaid on the data (black histogram). The (data--model)/error residuals are shown at the bottom of each panel (purple histogram), while the 68 and 95 per cent confidence intervals are shown as the dark and light shaded bands, respectively. The corrected zero-level is shown by the green dashed line, while the normalized continuum is shown as the long-dashed blue line. Tick marks above each spectrum show the locations of the \hef\,\textsc{i}$^{*}$ absorption (gray ticks) and the \het\,\textsc{i}$^{*}$ absorption (blue and red ticks indicate the $F=0.5$ and $F=1.5$ levels, respectively; cf. the same color scheme as in Figure~\ref{fig:grotrian}). The top right panel is a zoom in of the top left panel. The feature marked with a T corresponds to two telluric absorption features, comprising a weak feature at $\sim+60~{\rm km~s}^{-1}$ and the main component at $\sim+70~{\rm km~s}^{-1}$. The reduced chi-squared of the best-fitting model is $\chi^{2}/{\rm dof}=1.104$.\label{fig:modelfits}}
\end{figure*}

Interstellar absorption lines are traditionally modelled with a Voigt profile, which assumes that the gas is distributed as a Maxwellian. Given the high S/N data involved, we noticed that the profile is asymmetric about the line centre, and is therefore not well-modelled by a single Voigt profile; a two component Voigt profile model is also insufficient to fully describe the line profile, within the uncertainties of our very high S/N data.

We have therefore modelled the line shape using a linear spline with a fixed knot spacing of $3-4~{\rm km~s}^{-1}$ (roughly corresponding to the instrument resolution). The linear spline is then convolved with a Gaussian profile (the width of this Gaussian is a free parameter) to construct a smooth and continuous, arbitrary line profile that is positive-definite. This process is therefore a hybrid between Voigt profile fitting and the apparent optical depth method \citep{SavSem91}; we fit a smooth, arbitrary representation of the line profile shape, which is simultaneously represented by multiple absorption lines.

Our derivation follows a very similar procedure to that formulated by \citet{SavSem91}. The observed line profile is given by the convolution of the intrinsic line profile, $I(\lambda)$, with the instrument profile, $\phi(\sigma_{\rm inst})$. The intrinsic line profile is given by:
\begin{equation}
    I(\lambda)=I(\lambda)_{0}\,e^{-\tau(\lambda)}
\end{equation}
where $I(\lambda)_{0}$ is the continuum and $\tau(\lambda)=N\,\sigma(\lambda)$ is the optical depth profile, which consists of the total column density ($N$) and the absorption cross section, $\sigma(\lambda)$. We can express the cross section in terms of the velocity relative to the center of the profile:
\begin{equation}
    \sigma(v)=\frac{\pi\,e^{2}}{m_{\rm e}\,c^2}\,f\,\lambda_{0}\,c\,{\cal S}(v)
\end{equation}
where $e^{2}/m_{\rm e}\,c^2$ is the classical electron radius, $\lambda_{0}$ is the rest wavelength of the transition, $f$ is the corresponding oscillator strength, $c$ is the speed of light, and ${\cal S}(v)$ is a smooth, normalized spline function, which is the convolution of a linear spline, ${\cal L}(v)$, with a Gaussian of velocity width $\sigma_{v}$:
\begin{equation}
    {\cal S}(v) = {\cal L}(v)\,\circledast\,G(\sigma_{v}).
\end{equation}
To summarise, the free parameters of this function include the column density ($N$), the Gaussian convolution width ($\sigma_{v}$), and the line shape values at each spline knot of the linear spline, ${\cal L}(v)$. Note that the redshift of the line is not a free parameter of the function; it is fixed at a value that is close to the maximum optical depth, since the redshift is degenerate with changing the line profile weights at each spline knot. The model profile is generated on a sub-pixel scale of $1.5~{\rm m~s}^{-1}$ (corresponding to a sub-pixellation factor of 1000) and rebinned to the native pixel resolution ($1.5~{\rm km~s}^{-1}$) after the model is convolved with the instrument profile.

We stress that it is the combined information of the shape and strength of the absorption features that allow us to pin down the column density; the \emph{shape} of the line core is largely set by the weak $^{3}$\HeIs\ absorption, while the rest of the line shape is set by the stronger $^{4}$\HeIs\ absorption. The relative \emph{strength} of the lines then sets the \heiso\ ratio. So, even though the cores of the $^{4}$\HeIw\ line profiles \emph{appear} saturated, the absorption lines are fully resolved and the S/N is very high; there is sufficient information about the profile shape from just the \HeIw\ line profile to pin down the $^{4}$\HeIs\ column density and the \heiso\ ratio using all of the contributing absorption features that make up the complete line profile (i.e. the $^{4}$\HeIs\ transitions to the $J'=0$ and $J'=1,2$ levels, in combination with the $^{3}$\HeIs\ transitions to the $J'=1,2$ levels). The higher order optical/UV transitions are also important to determine the $^{4}$\HeIs\ column density, because they are weaker and are situated in the linear regime of the curve of growth.

However, given that the optical and near-infrared data were taken $\sim7$ years apart, the absorption profile might not be expected to have exactly the same shape at the two epochs, given the very high S/N of the data. We therefore model a separate spline to the UVES and CRIRES data, but all spline models are assumed have the same \het/\hef\ ratio. Finally, there are several telluric absorption lines that are located near the \HeIw\ absorption line; we model these telluric features as a Voigt profile with an additional damping term to account for collisional broadening.

Finally, the relative population of the $F=0.5$ and $F=1.5$ hyperfine levels of $^{3}$\HeIs\ is given by:
\begin{equation}
    \frac{n_{0.5}}{n_{1.5}} = \frac{g_{0.5}}{g_{1.5}}\exp(-T_{\star}/T_{s})
\end{equation}
where $g_{0.5}=2$ and $g_{1.5}=4$ are the level degeneracies, $T_{\star}=0.32~{\rm K}$, and $T_{s}$ is the spin temperature. Given that the spin temperature is expected to greatly exceed $T_{\star}$, the relative population of the $F=0.5$ and $F=1.5$ hyperfine levels is simply given by the ratio of the level degeneracies; as part of the fitting procedure, we appropriately weight the $^{3}$\HeIs\ transitions to account for the relative populations of the ground state. The data and the best-fitting model are shown in Figure~\ref{fig:modelfits}; the reduced chi-squared of this fit is $\chi^{2}/{\rm dof}=1.104$.

The best-fitting model corresponds to a total metastable $^{4}$\HeIs\ column density of $\log_{10\,}N(^{4}{\rm He\,\textsc{i}^{*}})/{\rm cm}^{-2}=13.6430\pm0.0031$ based solely on the \HeIw\ absorption in the CRIRES data. The optical data (based on a simultaneous fit to both \HeIs\,$\lambda3188$\,\AA\ and $\lambda3889$\,\AA) suggests a higher value of $\log_{10\,}N(^{4}{\rm He\,\textsc{i}^{*}})/{\rm cm}^{-2}=13.6972\pm0.0023$, which is statistically inconsistent with the CRIRES data. This difference may reflect a real change to the line profile depth between the two epochs of observation, and highlights the importance of recording as many absorption lines simultaneously; the weak higher order transitions are optically thin, and can be used to pin down the column density of $^{4}$\HeIs\ more reliably. Given the current data, the column density of \HeIs\ is decreasing at a rate of $-(2.29\pm0.16)\times10^{9}~{\rm atoms~cm}^{-2}~{\rm yr}^{-1}$; at this rate, the \HeIs\ absorption will be short-lived ($\sim20,000~{\rm years}$). This noticeable change between the two epochs could be due to the transverse motion of the cloud, or a reduction in the ionization parameter at the surface of the absorbing medium \citep{Liu2015}. Future optical data covering the weak \HeIs\,$\lambda3188$ absorption will help to pin down the time evolution of the strength and line-of-sight motion of this profile. For the present analysis, we do not use the optical absorption lines to determine the helium isotope ratio.

As mentioned earlier, we perform a direct fit to the total helium isotope ratio. The best-fitting value of the absorbing gas cloud towards \tet\ is:
\begin{equation}
    \log_{10}~^{3}{\rm He}/^{4}{\rm He} = -3.752\pm0.032
\end{equation}
or expressed as a linear quantity:
\begin{equation}\label{eqn:heisoratio}
    ^{3}{\rm He}/^{4}{\rm He} = (1.77\pm0.13)\times10^{-4}
\end{equation}
This corresponds to a seven per cent determination of the helium isotope ratio, and is currently limited only by the S/N. As a sanity check of this measurement, we use the $^{3}$\HeIs\ column density estimated from the equivalent width of the $^{3}$\HeIw\ absorption feature (Equation~\ref{eqn:ewcoldens}) in combination with the $^{4}$\HeIs\ column density based on the \textsc{alis} fits, to infer a helium isotope ratio of \heiso~$=(1.66\pm0.10)\times10^{-4}$. This estimate is consistent with the value based on our \textsc{alis} fits (Equation~\ref{eqn:heisoratio}); the difference between these two values of \heiso\ and their uncertainties, is due to the uncertain continuum placement when estimating the equivalent width (i.e. Equation~\ref{eqn:ewcoldens} does not include the uncertainty due to continuum placement). This highlights the benefit of simultaneously fitting the continuum and the absorption lines with \textsc{alis}: The continuum uncertainty is folded into all parameter values, and avoids introducing a systematic bias due to the manual placement of the continuum.

\begin{figure}
\includegraphics[width=\columnwidth]{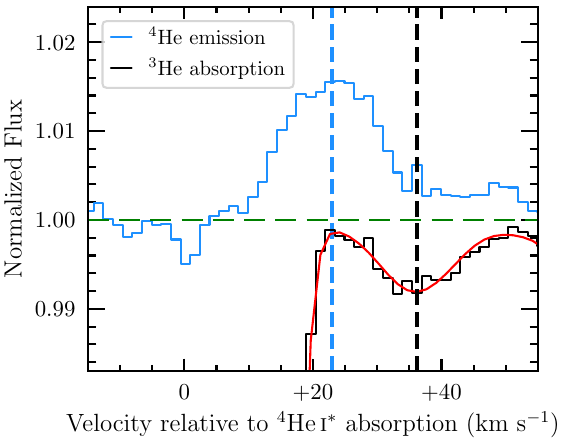}
\caption{Spectrum of \tet\ (black histogram) overlaid with the best-fitting model (red curve). The blue histogram shows the surrounding $^{4}$\HeIs\ emission that appears on the red wing of the absorption profile (cf. Figure~\ref{fig:2draw}); the emission profile has been offset by $+1$ (but is the same relative scale as the black histogram), to allow a close comparison to the absorption profile. The horizontal green dashed line simultaneously represents the normalized continuum of \tet, and the zero-level of the $^{4}$\HeIs\ emission line. The vertical blue and black dashed lines indicate the centroids of the $^{4}$\HeIs\ emission and $^{3}$\HeIs\ absorption, respectively.\label{fig:absemis}
}
\end{figure}

\subsection{Validation}
\label{sec:validate}

We performed several checks to validate the detection of $^{3}$\HeIs\ absorption towards \tet. First, we confirmed that there are no telluric absorption lines that are coincident with the locations of either the $^{3}$\HeIs\ or the $^{4}$\HeIs\ absorption. The only telluric lines nearby are those identified in Figure~\ref{fig:modelfits}.

As mentioned in Section~\ref{sec:obs}, to optimize the final combined S/N, we subtracted two raw frames at different nod positions before extraction. We also extracted spectra of \tet\ using the individual raw frames. The final combined S/N of the data based on this `alternative' reduction is $\sim600$ in the continuum, a factor of $\gtrsim2$ lower than using differencing. Nevertheless, we confirmed that the $^{3}$\HeIs\ absorption feature is present in the alternative reduction, and is therefore not an artifact of the frame differencing.

We also note that the faint $^{4}$\HeIs\ emission seen in Figure~\ref{fig:2draw} is not aligned with the $^{3}$\HeIs\ absorption feature (see Figure~\ref{fig:absemis}). Thus, the $^{3}$\HeIs\ absorption feature is not an artifact of subtracting frames at two different nod positions. Moreover, we note that the faint $^{4}$\HeIs\ emission feature peaks at $\sim+23~{\rm km~s}^{-1}$ relative to the absorption (or, $\sim+25~{\rm km~s}^{-1}$ in the heliocentric frame). The location and width of this emission feature is not easily attributable to any of the velocity structures previously identified in Orion \citep{Odell93}. Curiously, the peak of this emission feature is located at the point of minimum optical depth in the red wing of the $^{4}$\HeIs\ absorption feature. Thus, the emission feature may be caused by photons that are recombining at the face of the \HeIs\ absorbing medium. These photons escape from the cloud where the optical depth is lowest.

Finally, we consider the rare possibility that the $^{3}$\HeIs\ absorption feature is actually due to coincident $^{4}$\HeIs\ absorption that occurs at exactly the expected location and strength of the $^{3}$\HeIs\ absorption feature. In principle, this obscure possibility can be ruled out by searching for the corresponding $^{4}$\HeIs\ absorption in the optical/UV absorption lines, since the isotope shift is different for different transitions. We investigated this possibility for the UVES data currently available, but these data are not of high enough S/N to rule out the possibility of satellite $^{4}$\HeIs\ absorption. However, we note that the red wing of the \HeIs\,$\lambda3889$~\AA\ absorption feature is clean and featureless to the noise level of the current data, so the possibility of satellite $^{4}$\HeIs\ absorption can be ruled out with future data.

\section{Results}\label{sec:results}

\het\ has only been detected a handful of times, and never outside of the Milky Way; many of these detections come from observations or studies of Solar System objects, while the rest are based on observations of the \het$^{+}$ $8.7~{\rm GHz}$ line from \HII\ regions \citep{BalBan18}. In this section, we discuss how measurements of the helium isotope ratio can inform models of Galactic chemical enrichment and stellar nucleosynthesis. Observations of \heiso\ can also be used for cosmology and to study the physics of the early Universe.

\subsection{Galactic Chemical Evolution of \het}
\label{sec:vice}

To interpret our new determination of the Galactic helium isotope ratio, we performed a series of galactic chemical evolution (GCE) models using the Versatile Integrator for Chemical Evolution (\vice; \citealt{JohnsonWeinberg2020}).\footnote{\vice\ is available from \url{https://vice-astro.readthedocs.io}}
Our implementation closely follows the model described by \citet{Johnson2021}; we briefly summarise the key aspects of this model below, and refer the reader to \citet{Johnson2021} for further details.

There are two key motivations for using \vice\ in this work. First, the combination of an inside-out star formation history and radial migration has already been shown to reproduce many chemical properties and the abundance structure of the Milky Way using these \vice\ models \citep{Johnson2021}. Thus, to ensure this agreement is maintained, the only changes that we make to the \citet{Johnson2021} model are the \het\ and \hef\ yields, and the starting primordial composition. The second motivation for using \vice\ is its numerically-constrained model of radial migration which allows stellar populations to enrich distributions of radii as their orbits evolve. This is an important ingredient in GCE models of the Milky Way because stellar populations may move significant distances before nucleosynthetic events with long delay times occur; \citet{Johnson2021} discuss this effect at length for the production of iron by Type Ia supernovae. Since the dominant \het\ yield comes from low mass stars with long lifetimes (e.g. \citealt{Larson1974}, \citealt{MaederMeynet1989}), it is possible that similar processes may affect the distribution of \het\ in the Galaxy.

\vice\ models the Milky Way as a series of concentric rings of uniform width $\delta R_\text{gal}$ out to a radius of 20~kpc, and we retain the choice of $\delta R_\text{gal} = 100$~pc from \citet{Johnson2021}. The gas surface density of each ring is given by the gas mass in each ring, divided by the area of the ring, where the gas mass is determined by a balance of infall, outflow, star formation, and the gas returned from stars. Our models assume an inside-out star formation history (SFH), with a functional form:
\begin{equation}
\label{eqn:infall}
    \dot{\Sigma}_\star(t|R_{\rm gal}) \propto (1-e^{-t/\tau_{\rm rise}})\,e^{-t/\tau_{\rm sfh}}
\end{equation}
where $R_{\rm gal}$ represents the galactocentric radius at the center of each ring, $\tau_{\rm rise}=2~{\rm Gyr}$ (i.e. the star formation history peaks around a redshift $z\approx2.5$), and $\tau_{\rm sfh}$ is the e-folding timescale of the star formation history, which depends on $R_{\rm gal}$. The relationship between $\tau_{\rm sfh}$ and $R_{\rm gal}$ is based on a fit to the relationship between stellar surface density and age of a sample of Sa/Sb spiral galaxies (\citealt{Sanchez2020}; see Figure~3 of \citealt{Johnson2021}). We set the star formation rate to zero when $R_{\rm gal}>15.5~{\rm kpc}$. The star formation law we adopt is a piecewise power law with three intervals defined by the gas surface density; the intervals and power law indices are based on the aggregate observational data by \citet{Bigiel2010} and \citet{Leroy2013}, in combination with the theoretically motivated star formation laws presented by \citet{Krumholz2018}. Outflows are characterized by a mass-loading factor, $\eta=\dot{M}_{\rm out}/\dot{M}_{\star}$, where $\dot{M}_{\rm out}$ is the outflow rate, and $\dot{M}_{\star}$ is the star formation rate. The radial dependence of our outflow prescription ensures that the late-time abundance gradient of oxygen agrees with observations \citep{Weinberg2017}. We assume a \citet{Kroupa2001} stellar initial mass function (IMF).

Our model implements a radial migration of stars from their birth radius based on the outputs from the \texttt{h277} cosmological simulation \citep{Christensen2012,Zolotov2012,Loebman2012,Loebman2014,BrooksZolotov2014}. The only quantities used from this simulation are the galactocentric birth radius, birth time, and the final galactocentric radius (the radius at the end of the simulation) of each star particle. We assume that stars make a smooth, continuous migration from their birth radius to their final radius with a displacement proportional to the square root of the star's age. This functional form is similar to the assumption used by \citet{Frankel2020} to model the radial migration of stars due to angular momentum diffusion. We simulate our Milky Way models for 13.2~Gyr, which is set by the outputs of the \texttt{h277} simulation; given the age of the Universe ($13.801\pm0.024~{\rm Gyr}$; \citealt{Planck2018}), star formation begins in our Milky Way models $\sim0.6~{\rm Gyr}$ after the Big Bang ($z\approx9$).

\begin{figure*}
\includegraphics[width=\textwidth]{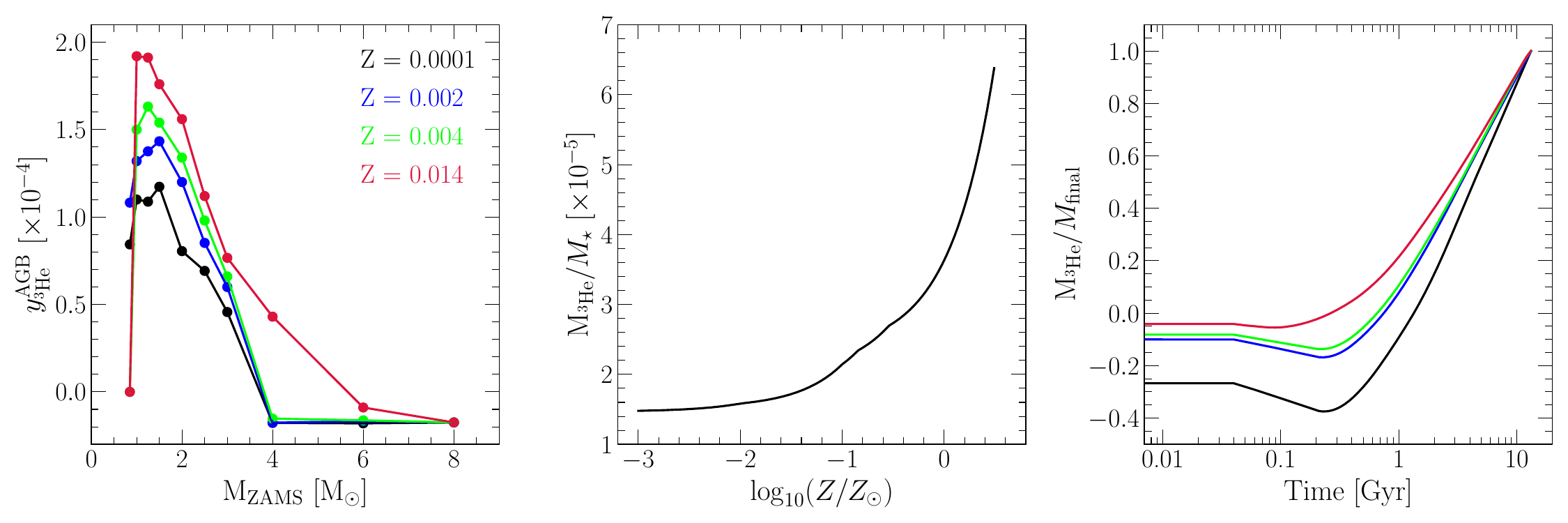}

\vspace{0.1cm}

\includegraphics[width=\textwidth]{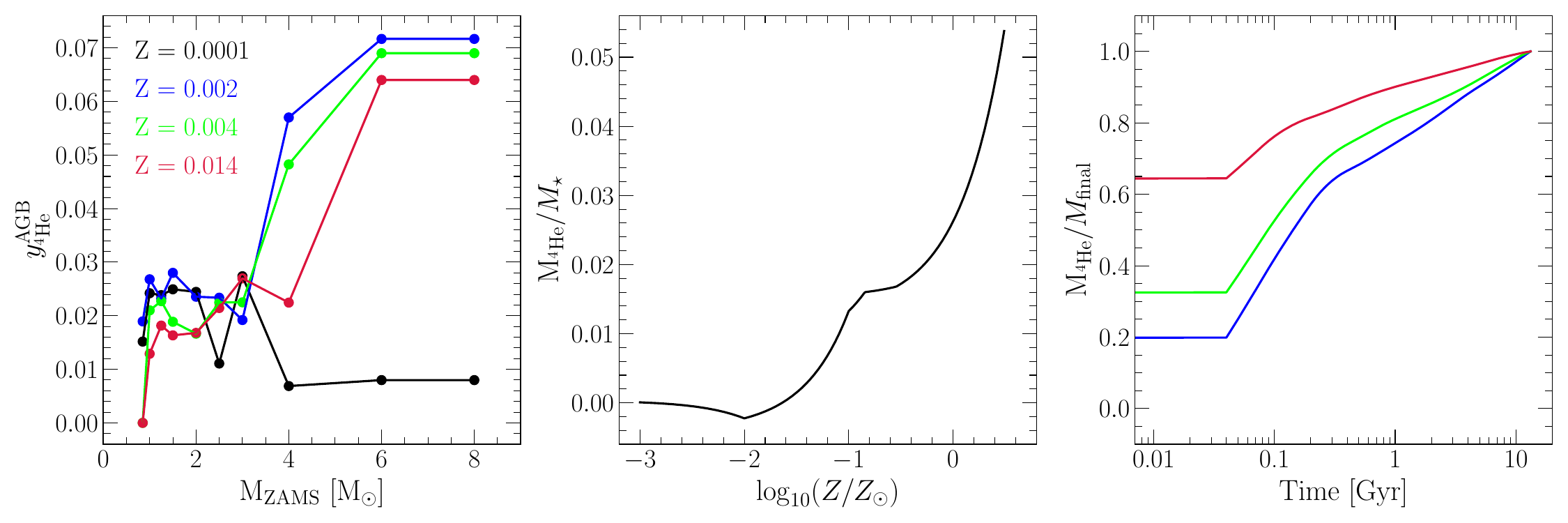}
\caption{The \het\ and \hef\ yields of AGB stars (top and bottom panels, respectively).
\textit{Left panels}: fractional yield as a function of the zero age main sequence (ZAMS) stellar mass. The yield of \het\ is dominated by the lowest mass stars, while the yield of \hef\ is dominated by intermediate and high mass stars.
\textit{Middle panels}: IMF integrated yield as a function of metallicity for a simple stellar population, normalized by the total stellar mass. The yields of both \het\ and \hef\ are greater for higher metallicity stellar populations.
\textit{Right panels}: fractional time evolution of \het\ and \hef\ for a simple stellar population. The \het\ yield is dominated by the lowest mass stars. As a result, radial migration is expected to be important for understanding the galactic chemical evolution of \heiso. We note that the $Z=0.0001$ curve is not shown in the bottom right panel for \hef, because extremely metal-poor massive stars net destroy more \hef\ than AGB stars net produce.
\label{fig:yields}
}
\end{figure*}

The key aspect of our models that differs from \citet{Johnson2021} is the choice of nucleosynthesis yields. Although \vice\ in its current version does not natively compute isotope-specific GCE models, the code base is built on a generic system of equations that make it easily extensible.\footnote{For the purposes of this paper, we simply replace the yields of a minor element (in our case, gold) with those of \het. Further details on \vice's implementation of enrichment can be found in its science documentation.} For stars that undergo core-collapse ($M>8\,{\rm M}_{\odot}$), \vice\ instantaneously deposits an IMF-weighted yield at the birth annulus. We calculate the IMF-weighted ($M>8\,{\rm M}_{\odot}$) metallicity-dependent net yields of \hef\ and \het\ using the nucleosynthesis calculations of \citet{LimongiChieffi2018};\footnote{Since the \citet{LimongiChieffi2018} yield tables only extend down to $13\,{\rm M}_{\odot}$, we perform a linear extrapolation over the mass range $8-13\,{\rm M}_{\odot}$.} we then linearly interpolate these IMF-weighted massive star yields over metallicity in \vice.\footnote{We use the `R' set with $v=0~{\rm km~s}^{-1}$, whereby the yields of all stars in the mass range $8\le M/{\rm M}_{\odot} \le25$ consist of both stellar winds and explosive nucleosynthesis, while the yields of stars with $M>25~{\rm M}_{\odot}$ only consist of stellar winds (i.e. stars with $M>25~{\rm M}_{\odot}$ are assumed to directly collapse to black holes).}
For example, at $Z={\rm Z_{\odot}}$, for every solar mass of stars formed in a ring, $0.065~{\rm M_{\odot}}$ of \hef\ is immediately returned to the ISM in that ring, of which $0.017~{\rm M_{\odot}}$ is freshly synthesized and $r\,Y_{\rm ISM} = 0.048~{\rm M_{\odot}}$ is recycled, where $Y_{\rm ISM}\simeq0.255$ is the ISM \hef\ mass fraction and $r = 0.19$ is the recycling fraction for $M > 8~{\rm M_{\odot}}$ stars assuming a \citet{Kroupa2001} IMF and that all stars with $M > 8~{\rm M_{\odot}}$ form a $1.44~{\rm M}_{\odot}$ compact remnant. The corresponding amount of hydrogen returned to the ISM is $0.076~{\rm M_{\rm \odot}}$. The amount of \het\ returned to the ISM is $5.9\times10^{-6}~{\rm M_{\odot}}$, which is lower than the recycled fraction ($7.4\times10^{-6}~{\rm M_{\odot}}$). Thus, the massive star yields that we employ act to reduce the \heiso\ ratio at solar metallicity. We note, however, that the \het\ and \hef\ yields of massive stars play a relatively minor role in the chemical evolution of \heiso; ignoring the yields of massive stars would result in an increase of the present day \heiso\ of Orion by $\sim4$ per cent.

For asymptotic giant branch (AGB) stars ($M<8\,{\rm M}_{\odot}$), we adopt the metallicity dependent net yields of \het\ and \hef\ reported by \citet{Lagarde2011,Lagarde2012}. Since the \citet{Lagarde2011,Lagarde2012} yields are only computed for $M\le6\,{\rm M}_{\odot}$, we assume that $8\,{\rm M}_{\odot}$ stars have the same net \hef\ yield as $6\,{\rm M}_{\odot}$ stars, while we assume the same \het\ yield for $8\,{\rm M}_{\odot}$ stars for all metallicities ($y_{\rm ^{3}He}^{\rm AGB} = -1.75\times10^{-5}$). We then linearly interpolate these yields over metallicity and stellar mass. We assume the \citet{Larson1974} metallicity-independent mass-lifetime relation for our calculations, and checked that our results were unchanged by using metallicity-dependent prescriptions \citep{HurleyPolsTout2000,Vincenzo2016}. In Figure~\ref{fig:yields}, we show the fractional \het\ and \hef\ net yields as a function of stellar mass and metallicity (left panels), the IMF-weighted yield as a function of metallicity (middle panels), and the gradual build up of \het\ and \hef\ for a simple stellar population (right panels). Taken together, these plots demonstrate that the lowest mass stars ($M\lesssim2\,{\rm M}_{\odot}$) are chiefly responsible for the production of \het, and this largely drives the evolution of the \heiso\ ratio. This is particularly true for solar metallicity AGB stars. As mentioned earlier in this Section, a key motivation of including the effects of radial migration is that the dominant \het\ yield comes from stars with the longest lifetimes, and therefore most likely to deposit their yield far from their birth galactocentric radius.\footnote{We have found that radial migration adds a small scatter of $\sim1-2$ per cent to the present-day \het/\hef\ values at all galactocentric radii; the overall GCE of \het/\hef\ is otherwise unchanged.}

We also update the primordial composition of \vice; using the \citet{Planck2018} determination of the baryon density (100\,\obhh~$= 2.242\pm0.014$), the latest value of the neutron lifetime ($\tau_{\rm n} = 879.4\pm0.6\,{\rm s}$; \citealt{PDG2020}), the $d(p,\gamma)^{3}{\rm He}$ rate reported by the Laboratory for Underground Nuclear Astrophysics (LUNA; \citealt{Mossa2020}), and assuming the Standard Model of particle physics and cosmology ($N_{\rm eff}=3.044$; \citealt{Mangano2005,deSalas2016,GrohsFuller2017,AkitaYamaguchi2020,EscuderoAbenza2020,Froustey2020,Bennett2020}), we determine the primordial helium isotope ratio to be (\het/\hef)$_{\rm p}=(1.257\pm0.017)\times10^{-4}$ (see \citealt{Pitrou21,Yeh2021}).

Before we use this model to infer the galactic chemical evolution of \heiso, it is worthwhile keeping in mind that some model ingredients are still missing. For example, the \vice\ Milky Way model has been shown to produce a broad range of [$\alpha$/Fe] at fixed [Fe/H] in the solar vicinity, but it does not produce a clear bimodality in [$\alpha$/Fe], in contrast with observations \citep{Vincenzo2021}. \citet{Johnson2021} conjecture that this discrepancy implies that the Milky Way's accretion and/or star formation history is less continuous than the model assumes, and changing this history could also alter the evolution of \heiso. One mechanism that can help to produce a stronger [$\alpha$/Fe] bimodality is the two-infall scenario \citep{Chiappini2001,Romano2010,Noguchi2018,Spitoni2019}. While we do not consider alternative (e.g. two-infall or bursty) star formation history in this work, we note that the observed \heiso\ ratio near the Orion Nebula is just $\sim 40$ per cent above the Standard Model primordial value; galactic chemical evolution has apparently altered the ISM ratio only moderately relative to its primordial value, a conclusion supported by the apparent similarity of primordial and ISM D/H ratios \citep{Linsky2006,Prodanovic2010}. \citet{Weinberg2017DH} argues that this weak evolution of D/H is a consequence of substantial ongoing dilution of the ISM by infall, with much of the material processed through stars ejected in outflows (see also, \citealt{Romano2006,Steigman2007}). Similar considerations would apply to \heiso.

\subsection{\het\ in the Milky Way}

\begin{figure*}
\includegraphics[width=0.95\textwidth]{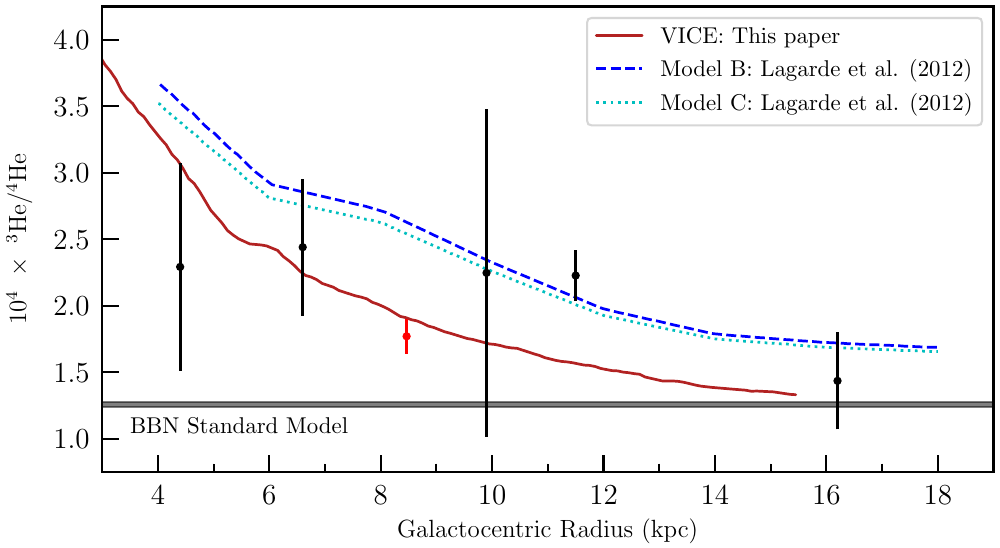}

\vspace{0.5cm}

\includegraphics[width=0.95\textwidth]{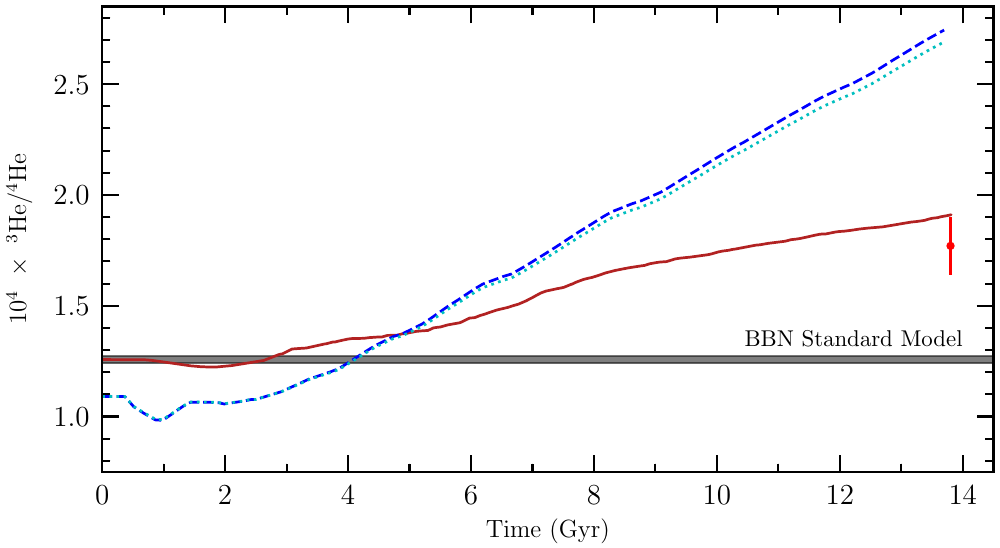}
\caption{Present-day radial profile (top panel) and time evolution at the galactocentric radius of the Orion Nebula (bottom panel) of the helium isotope ratio. Our \vice\ galactic chemical enrichment model is shown as the red curve. This model assumes that all low mass stars undergo the thermohaline instability and rotational mixing. Model C uses the same low mass star yields, but different input physics \citep{Lagarde2012}. Model B assumes that 96 per cent of stars undergo the thermohaline instability and rotational mixing, while the yields of the remaining stars adopt the `standard' stellar models (see \citealt{Lagarde2012} for further details about model B and C, and see Section~\ref{sec:vice} for details about our \vice\ models). Note that the \citet{Lagarde2012} models assume a different primordial composition; for a fairer comparison, the \citet{Lagarde2012} models should be increased by about $0.15~\times10^{-4}$. Our Orion Nebula \heiso\ determination is shown as the red symbol and error bar. In the top panel, we also overplot measures of the \het/H abundance of Galactic \HII\ regions \citep{BalBan18} converted to the helium isotope ratio using the estimated helium abundance of the \HII\ regions (see Equation 2 of \citealt{BalBan18}). The primordial \het/\hef\ value, assuming the Standard Model, is shown by the horizontal dark gray bands, (\het/\hef)$_{\rm p}=(1.257\pm0.017)\times10^{-4}$.\label{fig:galrad}
}
\end{figure*}

Early models of stellar nucleosynthesis indicated that low mass ($1-3~{\rm M}_{\odot}$) stars yield copious amounts of \het\ as part of the p-p chain while burning on the main sequence \citep{Iben67a,Iben67b,Rood72}. It was recognised soon after, in the context of GCE, that these stellar models may overpredict the amount of \het\ compared to the protosolar value (\citealt{Rood76}; see also, \citealt{TruCam71}). The stellar and GCE models were also at odds with the first observations of \het\ from \HII\ regions \citep{Rood79}. It was later recognized that the best determination of the \het/H abundance came from observations of structurally simple \HII\ regions \citep{Balser99HII, Bania02}. This pioneering work demonstrated that the \het\ abundance of the Milky Way is mostly the same at different locations. Moreover, the value derived by \citet{Bania02} is of similar magnitude to the protosolar value, indicating that the \het\ abundance did not change significantly during the past 4.5~Gyr of Galactic evolution.

It became clear that models of stellar nucleosynthesis were overproducing \het, and missing a critical ingredient, leading to the so-called \het\ problem. \citet{SacBoo99} proposed that additional mixing can help destroy \het, and can alleviate the discrepancy between the protosolar \het\ value and GCE models \citep{Palla00,Chiappini02}. A possible mechanism for the additional mixing --- the thermohaline instability --- was identified by \citet{ChaZah07a} as a physical mechanism to resolve the \het\ problem (see also, \citealt{ChaZah07b}). Models of GCE combined with a grid of stellar models that employ the thermohaline instability and rotational mixing \citep{Lagarde2011,Lagarde2012} were found to produce remarkable agreement with the protosolar and present day abundance of \het\ in the Milky Way, as well as the radial profile of \het\ from \HII\ regions \citep{BalBan18}.

In Figure~\ref{fig:galrad}, we show the present day radial \heiso\ profile (top panel) and the time evolution of \heiso\ at the galactocentric distance of the Orion Nebula (bottom panel). The red curve shows the results of our \vice\ models (see Section~\ref{sec:vice}), and the red symbol and error bar represents our determination of the \heiso\ ratio of the Orion Nebula. We also overplot the results of two GCE models presented by \citet{Lagarde2012}: Model B (blue dashed curve) assumes that 4 per cent of low mass stars obey the `standard' stellar evolution models, while the remaining 96 per cent undergo the thermohaline instability and rotational mixing. Model C (cyan dotted curve) assumes that all low mass stars undergo the thermohaline instability and rotational mixing. Note, the initial primordial composition assumed by \citet{Lagarde2012} is lower than the current Standard Model value; correcting for this offset would result in Model B and C being shifted to higher \heiso\ values. The black symbols with error bars illustrate the latest \het/H measures from \citep{BalBan18}, converted to a helium isotope ratio using the conversion (see their Equation~2):
\begin{equation}
    ^{4}{\rm He/H} = 0.105 - 1.75\times10^{-3}~R_{\rm gal}.
\end{equation}
The primordial helium isotope ratio, (\het/\hef)$_{\rm p}=(1.257\pm0.017)\times10^{-4}$, is shown by the horizontal dark gray bands in Figure~\ref{fig:galrad}. Our \vice\ models are in good agreement with our determination of the Orion helium isotope ratio and measures of the \het/H abundance of Galactic \HII\ regions. The \citet{Lagarde2012} chemical evolution models are vertically offset from our \vice\ GCE models and does not provide as good fit to our determination of the Orion helium isotope ratio.

To diagnose this offset, we explored the sensitivity of various model parameters to the chemical evolution of \heiso. The conclusion of these tests is that the relative importance of outflows and inflows has the strongest impact on the chemical evolution of \heiso. With strong outflows, significant infall of primordial gas is necessary to sustain ongoing star formation, and a larger portion of the ISM is made up of unprocessed material. Models that have weaker outflows retain more freshly synthesized \het\ in the ISM, leading to a higher \heiso\ ratio. The strength of outflows is currently an uncertain parameter in GCE models. 
Our \vice\ models employ strong outflows in order to reproduce the observed Milky Way oxygen gradient given our assumed IMF averaged yield of oxygen $y_O=0.015$ ($1.5 M_\odot$ of oxygen produced per $100 M_\odot$ of star formation). Some other GCE models omit outflows entirely
(e.g., \citealt{Spitoni2019,Spitoni2021}), and can still produce an acceptable disc if this oxygen yield is much lower, perhaps because of a steeper IMF or more extensive black hole formation \citep{Griffith2021}. Like (D/H), the \heiso\ ratio may offer important constraints on the strength of outflows because much of the ISM \het\ and \hef\ are primordial in origin and thus sourced by ongoing accretion (see Section~\ref{sec:impcosmo} below). As a side note, we found that the stellar IMF and the mass-lifetime relation of stars have a negligible impact on our results.

\subsection{Implications for Cosmology}
\label{sec:impcosmo}

Since our determination of the helium isotope ratio is obtained from a relatively metal-enriched environment, the composition that we measure does not directly reflect the primordial \heiso\ ratio. However, we consider two approaches below that allow us to infer the most likely primordial \heiso\ ratio, based on the currently available data: An empirical measure, and a model-dependent determination.

\begin{figure*}
\includegraphics[width=\textwidth]{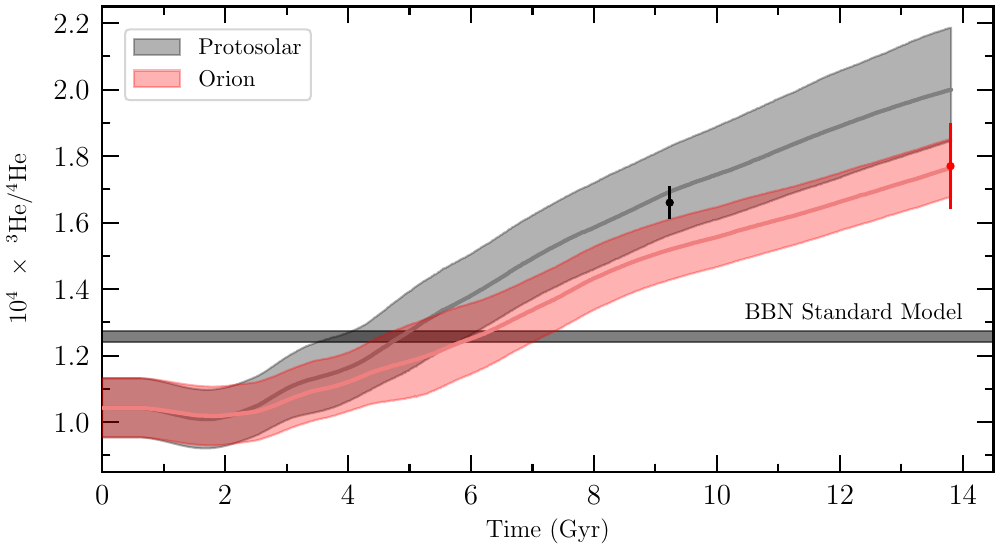}
\caption{The time evolution of the Milky Way \heiso\ ratio according to our \vice\ Galactic Chemical Evolution models (gray and red bands). The gray band is for a galactocentric radius of $7.3\pm0.6\,{\rm kpc}$ (corresponding to the birth galactocentric radius of the Sun), while the red band  is for $8.47\,{\rm kpc}$ (corresponding to the galactocentric radius of Orion). The black and red symbols with error bars are the protosolar measurement and present day Orion measurement of \heiso, respectively. The width of each band represents the $1\sigma$ uncertainty of the primordial \heiso\ ratio. Note that the width of the gray band also includes the uncertainty of the birth galactocentric radius of the Sun. The horizontal dark gray band represents the Standard Model value of the primordial \heiso\ ratio.
\label{fig:timeevol}
}
\end{figure*}

We first consider an empirical assessment of the primordial \het/H ratio, based on a similar approach as that outlined by \citet{Yang1984}. We start by converting our measure of the helium isotope ratio to a \het\ abundance, using the measured \hef/H ratio of the Orion nebula (${\rm ^{4}He/H}=0.0913\pm0.0042$; \citealt{MesaDelgado2012}); we estimate that the \het\ abundance of Orion is $^{3}{\rm He/H}=(1.62\pm0.14)\times10^{-5}$. We then note the following inequality:
\begin{equation}
    \bigg(\frac{^{3}{\rm He+D}}{\rm H}\bigg)_{\rm p} \le \bigg(\frac{^{3}{\rm He+D}}{\rm H}\bigg)_{\dagger}
\end{equation}
where subscript `p' refers to the primordial value, and the subscript `$\dagger$' refers to the Orion value. This inequality holds because: (1) Essentially all D is burnt into \het\ at temperatures $\gtrsim6\times10^{5}~{\rm K}$; and (2) Models of stellar nucleosynthesis indicate that the dominant \het\ yield comes from low mass stars, which are net producers of \het. Rearranging this inequality, we can solve for the primordial \het/H\ ratio, given the measured values of the primordial deuterium abundance $({\rm D/H})_{\rm p} = (2.527\pm0.030)\times10^{-5}$ \citep{Cooke18}, and the interstellar deuterium abundance ${\rm (D/H)}_{\dagger}=(2.0\pm0.1)\times10^{-5}$ \citep{Prodanovic2010}.\footnote{The Milky Way D/H abundances used by \citet{Prodanovic2010} are all within $\lesssim1\,{\rm kpc}$ of the Sun \citep{Linsky2006}. Therefore, we assume that this D/H abundance is similar to that of the Orion Nebula.} Combined with our estimate of the Orion \het/H ratio, we find:
\begin{equation}
    (^{3}{\rm He/H})_{\rm p} \le (1.09\pm0.18)\times10^{-5}
\end{equation}
where the uncertainties of the interstellar helium and deuterium abundances contribute almost equally to the total quoted uncertainty. This limit is in agreement with the Standard Model value, $(^{3}{\rm He/H})_{\rm p}=(1.039\pm0.014)\times10^{-5}$ \citep{Pitrou21,Yeh2021}.\footnote{One of the assumptions underlying this estimate is that the high dispersion seen in local D/H measures of the ISM is due entirely to the preferential depletion of deuterium onto dust grains  \citep{Linsky2006,Prodanovic2010}. However, it has since been discovered that there are intrinsic variations in the degree of metal enrichment of the ISM on small physical scales ($\sim$tens of pc; \citealt{deCia2021}). If we instead assume a local ISM D/H abundance based on the value measured from within the Local Bubble, ${\rm (D/H)}_{\dagger}=(1.56\pm0.04)\times10^{-5}$ (i.e. based on systems with $\log~N({\rm H\,\textsc{i}})/{\rm cm}^{-2} \le19.2$; \citealt{Linsky2006}), the limit on the primordial \het\ abundance becomes $(^{3}{\rm He/H})_{\rm p} \le (0.65\pm0.15)\times10^{-5}$, which is $2.6\sigma$ below the Standard Model value.}

In addition to our empirical limit on the primordial \het\ abundance, we also consider a model-dependent estimate of the primordial helium isotope ratio, based on our best available understanding of stellar nucleosynthesis and chemical evolution. We use our \vice\ GCE models to infer the best-fitting primordial \heiso, given two time-separated determinations of the helium isotope ratio of the Milky Way: The first measure is the protosolar value, which is based on a measurement of Jupiter's atmosphere by the Galileo Probe Mass Spectrometer \citep{Mah1998} (i.e. a snapshot of the Milky Way helium isotope ratio some $4.5~{\rm Gyr}$ ago). The second determination that we use is the present day Milky Way value reported in this paper.
These two values are remarkably comparable, and are only slightly elevated above the Standard Model value.
Given the very gradual change to the primordial \hef/H abundance over time ($\sim11$ per cent), this suggests that the build up of \het, even in a galaxy such as the Milky Way, is relatively gradual, probably because infall continually drives it back toward the primordial value. Thus, even in a chemically evolved galaxy, such as the Milky Way, we can still estimate the primordial \heiso\ ratio because: (1) The build up of \het\ is gradual over time; and (2) We have two time-separated measures of the \heiso\ ratio covering one-third of the Milky Way's age.

GCE models indicate that the \heiso\ value depends on both galactocentric radius and the amount of time that has passed since the start of chemical evolution. The present day measurement of the \heiso\ ratio towards \tet\ is based on a gas cloud in the vicinity of the Orion Nebula, estimated to be at a galactocentric radius of $\sim8.5\,{\rm kpc}$. The current age of the Universe is $13.801\pm0.024\,{\rm Gyr}$ \citep{Planck2018}, which in our case corresponds to the final output of the \vice\ models. The protosolar \heiso\ measurement reflects the Milky Way value at the birth of the Solar System $4.5682\,{\rm Gyr}$ ago \citep{BouvierWadhwa10}. The birth galactocentric radius of the Sun is estimated to be somewhat closer to the galactic center than the present distance, due to a combination of radial heating and angular momentum diffusion \citep{Minchev2018,Frankel2020}; in our analysis, we use the semi-empirical result of \citet[][$R_{\rm gal,\odot}=7.3\pm0.6\,{\rm kpc}$]{Minchev2018}.\footnote{We also repeated our analysis with the recent result by \citet[][$R_{\rm gal,\odot}=7.8\pm0.6\,{\rm kpc}$]{Frankel2020}, and the inferred (\heiso)$_{\rm p}$ increased by $\lesssim0.6\sigma$.}

\begin{figure*}
\includegraphics[width=\textwidth]{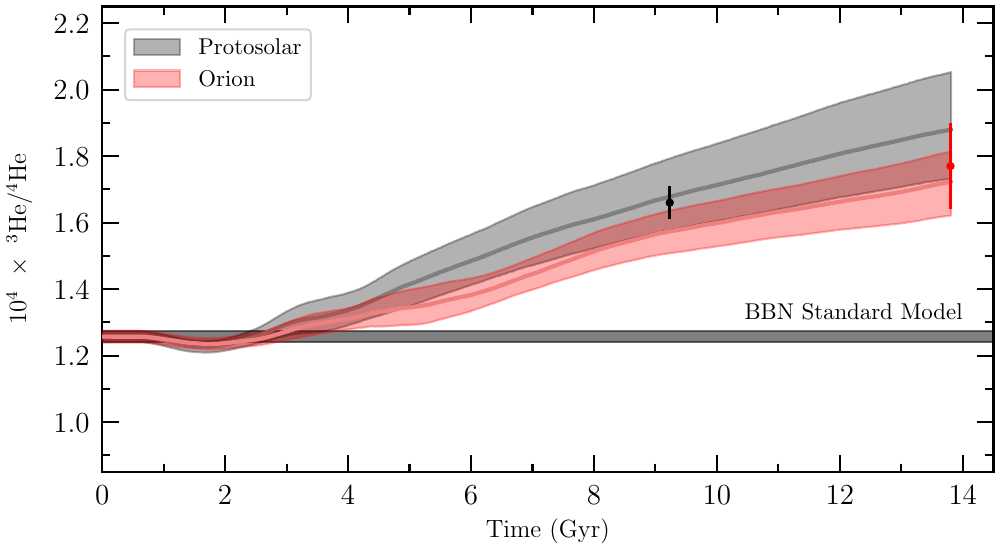}
\caption{Same as Figure~\ref{fig:timeevol}, but the \vice\ model shown here assumes the Standard Model \heiso\ ratio and allows the strength of the outflow prescription to vary as a free parameter. The best-fit models scale the default
outflow mass loading efficiencies upward by a factor $f_s=1.7$.
\label{fig:timeevol_outflows}
}
\end{figure*}

We generated a suite of \vice\ models covering a grid of primordial \heiso\ values. For each \heiso\ value, we take an average of 16 \vice\ simulations to minimize the post-BBN scatter of \heiso\ values due to radial migration (this scatter is of order $1-2$ per cent). We include this scatter as part of the model uncertainty, even though this is subdominant compared to the measurement errors. To determine the most likely value of the primordial \heiso\ ratio, we linearly interpolated over our averaged grid of \vice\ models, and then conducted a Markov Chain Monte Carlo analysis using the \textsc{emcee} software \citep{emcee}. We assume a uniform prior on the \heiso\ ratio, and a Gaussian prior on the birth galactocentric radius of the Sun, as described above. We note that, aside from the assumptions stated above, the only free parameter of this model is the primordial \heiso\ value. Given our \vice\ model, the best-fit value and 68 per cent confidence interval of the primordial \heiso\ ratio is:
\begin{equation}
\label{eqn:primheiso}
    (^{3}{\rm He}/^{4}{\rm He})_{\rm p} = (1.043\pm0.089)\times10^{-4}
\end{equation}
We remind the reader that this value is model dependent, and the error term does not include the (presently unknown) uncertainty associated with the nucleosynthetic yields and the GCE model. However, given that this determination is within $2.4\sigma$ of the Standard Model value, (\het/\hef)$_{\rm p}=(1.257\pm0.017)\times10^{-4}$ without accounting for the unknown model error, we consider our result to be in agreement with the Standard Model to within $\sim2\sigma$. Given that our model has a single free parameter --- the primordial \heiso\ ratio --- it is remarkable that our GCE model and yields can simultaneously reproduce the protosolar and present day values of the Galactic \heiso\ ratio without any tuning. The time evolution of our best-fit model is shown in Figure~\ref{fig:timeevol}, where the gray and red shaded regions are for the protosolar and present day \heiso, respectively.

We also consider an alternative model where we fix the primordial \heiso\ to the Standard Model value, and tune the strength of the outflow prescription to match the currently available data. The relative contributions of outflows (removing freshly synthesised \het\ and \hef) and inflows (of primordial material) is the dominant factor that sets the GCE of \heiso. In our model the gas surface density is determined by the star formation rate and the empirically motivated star formation law \citep{Johnson2021}, so a higher outflow also implies a higher inflow to replenish the gas supply. For our alternative model, we scale the strength of our outflow prescription uniformly at all radii, so that $\eta_{\rm new}=f_{s}\,\eta$. As described in Section~\ref{sec:vice}, the radial dependence and strength of the outflow prescription is defined in \vice\ to match the late-time abundance gradient of oxygen to observations \citep{Weinberg2017}. Thus, by rescaling the strength of outflows with the above prescription, we are altering the abundance distribution of oxygen in our models. In order to reconcile these rescaled models with the late-time abundance gradient of oxygen, the total oxygen yield would need to be comparably scaled (i.e. stronger outflows require a higher oxygen yield to match the data).

We generated a grid of \vice\ models where the outflows are uniformly scaled by $f_{s}$, and sampled this grid using an MCMC analysis similar to that described earlier. We determine a best-fitting value of the outflow scale factor, $f_{s}=1.70_{-0.25}^{+0.31}$, which is able to simultaneously reproduce the protosolar and present-day value of the \heiso\ ratio, assuming the Standard Model primordial \heiso\ value. The time evolution of the \heiso\ ratio of the Milky Way for this alternative model is shown in Figure~\ref{fig:timeevol_outflows}. Our models presume that gas ejected in outflows has the same chemical composition as the ambient ISM. We have not examined models where the winds are preferentially composed of massive star ejecta, which might arise if core collapse supernovae are the primary wind drivers, but we expect that they would require higher outflow mass-loading because more of the $^3$He produced by low mass stars would be retained. Finally, we note that the \textit{combination} of D/H and \heiso\ may provide strong constraints on the strength of outflows and thus indirectly constrain the scale of the yields, since: (1) both D and \het\ are primordial, and for the Standard Model, D/H and \heiso\ are known; (2) D is completely burnt into \het; and (3) \het\ is net produced by low mass stars.

\subsection{Future prospects}

Helium-3 is rarely detected in any environment, and never outside of the Milky Way. The new approach developed in this paper offers a precise measure of the helium isotope ratio, and is currently limited only by the S/N of the observations. Future measurements along different sightlines toward the Orion Nebula may further improve this determination, and test the consistency of this technique. Several sightlines suitable for carrying out this test are already known towards the brightest stars in Orion \citep{Odell93,Oud97}. Going further, \HeIs\ absorption is also detected towards $\zeta$~Oph \citep{GalKre12} and several stars in the cluster NGC 6611 \citep{Evans05} in the Milky Way, that would permit a measure of the radial \heiso\ abundance gradient (see Figure~\ref{fig:galrad}) and obtain further tests of our GCE model and pin down the inferred primordial abundance.

Another critical goal is to measure \heiso\ in more metal-poor environments. Fortuitously, the Small and Large Magellanic Clouds offer the perfect laboratory for carrying out such an exploration. Suitable stars in the Magellanic Clouds are bright enough ($m_{\rm J}\simeq12-15$) to confidently detect ($5\sigma$) a $^{3}$\HeIs\ absorption line with an equivalent width, $EW=3.6~{\rm m\AA}$ (i.e. identical to the feature towards \tet) in $\sim50$~hours with the upgraded VLT+CRIRES. This exposure time could be reduced for sightlines with a higher \HeIs\ column density. While challenging, this experiment is achievable with current facilities if suitable targets can be identified.

Finally, we note that it may be possible to detect \het\ in absorption against gamma-ray burst or quasar spectra; strong \HeIs\ absorption has already been detected towards GRB\,140506A at redshift $z=0.889$ \citep{Fynbo14}, but to reach the requisite S/N, a detection of $^{3}$\HeIs\ may require a significant investment of telescope time and rapid response observations. Because GRBs fade on short timescales ($\lesssim1~{\rm day}$), this may only be possible with the next generation of $30+$\,m telescopes.

\section{Conclusions}\label{sec:conc}
We have presented the first detection of metastable $^{3}$\HeIs\ absorption using the recently upgraded CRIRES spectrograph at the VLT, as part of science verification. The absorption, which occurs along the line-of-sight to \tet\ in the Orion Nebula, is detected at high confidence ($>13\sigma$), and has allowed us to directly measure the helium isotope ratio for the first time beyond the Local Interstellar Cloud. Our conclusions are summarized as follows:

\smallskip

(i) The helium isotope ratio in the vicinity of the Orion Nebula is found to be $^{3}{\rm He}/^{4}{\rm He}=(1.77\pm0.13)\times10^{-4}$, which is roughly $\sim40$ per cent higher than the primordial abundance assuming the Standard Model of particle physics and cosmology.

\smallskip

(ii) We calculated a suite of galactic chemical evolution models with \vice\ using the best available chemical yields of low mass stars that undergo the thermohaline instability and rotational mixing. This model confirms previous calculations that the \heiso\ ratio decreases with increasing galactocentric radius. Our model, which reproduces many chemical properties and the abundance structure of the Milky Way, is in good agreement with our measurement of \heiso.

\smallskip

(iii) We use these models to perform a joint fit to the present day (Orion Nebula) \heiso\ value and the protosolar value (a snapshot of the Milky Way $\sim4.5~{\rm Gyr}$ ago), allowing only the primordial helium isotope ratio to vary. Our models can reproduce both time-separated measurements if the primordial helium isotope ratio is $(^{3}{\rm He}/^{4}{\rm He})_{\rm p} = (1.043\pm0.089)\times10^{-4}$,
which agrees with the Standard Model value to within $\sim2\sigma$. We remind the reader that the quoted confidence interval does not include the model and yield uncertainty. We also report a more conservative, empirical limit on the primordial \het\ abundance
$(^{3}{\rm He/H})_{\rm p} \le (1.09\pm0.18)\times10^{-5}$, which is based on the measured \hef/H ratio of Orion, and the amount of primordial deuterium that has been burnt into \het.

\smallskip

(iv) As an alternative to this analysis, we can reproduce the protosolar and present-day values of \heiso\ if we assume the Standard Model primordial \heiso\ value and scale the strength of outflows in our \vice\ models. However, the strength of the outflows would need to be uniformly increased by a scaling factor $f_{s}=1.70_{-0.25}^{+0.31}$, which would in turn require comparable increase in oxygen and iron yields to retain the empirical successes of this model found by \citet{Johnson2021}. Our measured \heiso\ ratio offers a stringent test for Milky Way chemical evolution models.

\smallskip

Detecting $^{3}$\HeIs\ absorption is challenging due to the rarity of metastable helium absorbers, and the high S/N required to secure a confident detection of a weak absorption feature. Although the Milky Way is not the ideal environment to secure an estimate of the primordial abundance, future measurements of \heiso\ in the Milky Way will allow us to better understand the galactic chemical evolution of \het. Furthermore, if suitable sightlines can be found towards stars in nearby star-forming dwarf galaxies, or along the line-of-sight to a low redshift gamma-ray burst (e.g. \citealt{Fynbo14}), this approach may offer a reliable technique to pin down the primordial helium isotope ratio, and thereby test the Standard Model of particle physics and cosmology in a new way. However, such an ambitious goal may have to wait until the forthcoming generation of telescopes with $30+\,{\rm m}$ aperture. Observations of local dwarf galaxies will require a high contrast between the stellar and nebula emission, to ensure that the surrounding \HeIs\ emission does not contribute significantly to the noise in the vicinity of the weak $^{3}$\HeIs\ absorption line. Observations of gamma-ray bursts will require that the GRB: (1) explodes in a metal-poor environment; (2) is at sufficiently low redshift ($z\lesssim0.66$) so that the \HeIw\ absorption line can still be detected with future facilities \citep{Marconi2021}; and (3) is sufficiently bright for a long enough time that the required S/N can be achieved.

Finally, we point out that \emph{three} primordial abundance measurements --- D/H, \hef/H, and \het/\hef\ --- all agree with the Standard Model values to within $\sim20$ per cent or much better. This is in stark contrast to the observationally inferred primordial $^{7}$Li/H abundance, which disagrees with the Standard Model value by $\sim350$ per cent (see the review by \citealt{Fields11}). It is therefore becoming increasingly difficult to explain this discrepancy --- dubbed the Cosmic Lithium Problem --- with physics beyond the Standard Model, without breaking the remarkable simultaneous agreement of the primordial D/H, \hef/H and \het/\hef\ ratios with the Standard Model of particle physics and cosmology.

\begin{acknowledgments}
We thank an anonymous referee for their review of our manuscript, and for the many helpful comments offered.
This paper is based on observations collected at the European Organisation for Astronomical Research in the Southern Hemisphere, Chile (VLT program IDs: 107.22U1.001, 194.C-0833). We are most grateful to the staff astronomers at the VLT for their assistance with the observations.
During this work, RJC was supported by a
Royal Society University Research Fellowship.
RJC acknowledges support from STFC (ST/T000244/1).
JWJ and DHW were supported by NSF grant AST-1909841.
LW acknowledges support from Fondazione Cariplo (grant number 2018-2329).
MTM acknowledges the support of the Australian Research Council through Future Fellowship grant FT180100194.
This research has made use of NASA's Astrophysics Data System.
\end{acknowledgments}

\vspace{5mm}
\facilities{VLT(CRIRES and UVES)}


\software{ALIS \citep{Cooke2014},
          astropy \citep{Astropy2013,Astropy2018},
          emcee \citep{emcee},
          matplotlib \citep{matplotlib},
          numpy \citep{numpy},
          PypeIt \citep{PypeIt},
          scipy \citep{scipy},
          sympy \citep{SymPy},
          VICE \citep{Johnson2021}.
          }


\bibliography{mnbib}{}
\bibliographystyle{aasjournal}



\end{document}